\documentclass[a4paper,12pt]{article}
\usepackage{amsmath,subfigure}
\usepackage{amssymb,graphicx}

\newcommand{\kap}{\varkappa}

\begin{document}

\title{Relativistic theory of inverse beta-decay
of polarized neutron in strong magnetic field}

\author{S.Shinkevich\thanks{E-mail: shinkevich@gmail.com},
 A.Studenikin\thanks{E-mail: studenik@srd.sinp.msu.ru}}
\date{}
\maketitle

\begin{center}
{\em Department of Theoretical Physics, Moscow State University,
119992, Moscow, Russia}
\end{center}

\begin{abstract}
The relativistic theory of the inverse beta-decay of polarized
neutron, $\nu _{e} + n \rightarrow p + e ^{-}$, in strong magnetic
field is developed. For the proton wave function we use the exact
solution of the Dirac equation in the magnetic filed that enables
us to account exactly for effects of the proton momentum
quantization in the magnetic field and also for the proton recoil
motion. The effect of nucleons anomalous magnetic moments in
strong magnetic fields is also discussed. We examine the cross
section for different energies and directions of propagation of
the initial neutrino accounting for neutrons polarization.  It is
shown that in the super-strong magnetic field the totally
polarized neutron matter is transparent for neutrinos propagating
antiparallel to the direction of polarization. The developed
relativistic approach can be used for calculations of cross
sections of the other URCA processes in strong magnetic fields.
\end{abstract}

\rightline {PACS No 23.40.Bw}

\section{Introduction}
 It is by now widely recognized that strong magnetic
fields can be a significant factor relevant to diverse
astrophysical and cosmological environments. The presence of
strongest magnetic fields in proto-neutron stars and pulsars is
well established. The surface magnetic fields for many
radio-pulsars, that can be estimated by the observed synchrotron
radiation, are of the order of $B\sim 10^{12} - 10^{14} G$. There
are also so-called magnetars \cite{ThoDun93-96,BarHar98} whose
surface magnetic fields are two or three orders of magnitude
higher.

Although the internal structure of the magnetic field of a neutron
star is controversial, its strength can be estimated \cite{Land67}
as a limit imposed by the requirement that the total energy of the
star, including gravitational, electromagnetic, and thermal
components, must be negative (the scalar virial theorem), so that
the star is a bounded system. The scalar virial theorem sets the
upper limit on the internal neutron star magnetic field on the
level of $B\sim 10^{18} G$ \cite{LaiSha91}. The same estimation
for the internal field one obtains if the magnetic field flux is
supposed to be the trapped primordial flux.

Very strong magnetic fields are also supposed to exist in the
early Universe (see e.g.\cite{GraRub01}). Such fields can
influence the primordial nucleosynthesis
\cite{Gre69,MatOCon70,GraRub96} and affect the rate of  $^{4}{He}$
production.

Under the influence of strong magnetic fields the direct URCA
processes like
\begin{align} \label{URCA}
  &n\rightarrow p + e +\bar\nu_e,\\
  &\nu_e + n\leftrightharpoons e+p,\\
  &p+\bar\nu_e \leftrightharpoons n+e^{+}
\end{align}
can be modified. These reactions play important role in the
neutron star evolution so that the presence of strong magnetic
fields significantly change the star cooling rate
\cite{LeiPer98,ArrLai99,Goy98,GvoOgn99,LaiShap91}. It is worth to
be noted here also a recent study of neutrino processes (2) and
(3) in strong magnetic fields of the order $10^{16} G$ and
implication for supernova dynamics \cite{HuaQian04_2}.

The direct URCA processes have gained a lot of attention because
of the asymmetry in the neutrino emission, which can arise in the
presence of strong magnetic fields. Various authors have argued
that asymmetric neutrino emission during the first seconds after
the massive star collapse could provide explanations for the
observed pulsar velocities. A lot of different mechanisms for the
asymmetry in the neutrino emission from a pulsar has been studied
previously (see e.g.
\cite{Chu84,DorRodTer84-85,KusSeg96,AkhLanSci97,NunSemSmiVal97,BisKog97,Rou97,LaiQia98,Stu88}).
For more complete references on the neutrino mechanisms of the
pulsar kicks see the review paper \cite{BhaPal02,Kusenko04_09}.

It is worth to be noted here that the angular dependence of the
neutrino emission in URCA processes was first considered for the
neutron beta-decay neutrinos in \cite{Kor64,TerLysKor65}. In these
papers the probability of the polarized neutron beta-decay in the
presence of a magnetic field was derived, as well as the asymmetry
in the neutrino emission was studied for the first time. In the
two well known papers, refs. \cite{MatOCon69} and \cite{Fas69},
the results of \cite{Kor64,TerLysKor65} for the neutron decay rate
in a magnetic field were re-derived, however there were no
discussion on the asymmetry in neutrino emission in refs.
\cite{MatOCon69,Fas69}.

The neutron beta-decay have been studied in different
electromagnetic field configurations. The first attempts to
consider the beta-decay in the field of an electromagnetic wave
have been undertaken in \cite {Gaz65} and \cite{Bar74}. However,
the final result of  \cite{Gaz65} is very complicated and is not
accessible for any numerical analysis, whereas the result of
\cite{Bar74} for the field influence on the decay rate is far
overestimated. In \cite{TerRodZhuStu78_80} we have considered the
probability of the polarized neutron beta-decay in the
superposition of a magnetic field and a field of an
electromagnetic wave (the so-called Redmond field configuration)
and have confirmed the results of \cite{Kor64,TerLysKor65} for the
decay probability in the magnetic field and also got the
probability in the presence of an electromagnetic wave field. The
relativistic theory of the beta-decay of the neutron (accounting
for the proton recoil motion) in the strong magnetic field has
been developed in \cite{Stu89}. Many important technical details
of the calculations, also useful for the studies performed in the
present paper, can be found in \cite{TerRodZhuStu78_80}. The rates
of the two inverse processes in eqs. (3) and (4) in the presence
of a magnetic field has been derived in \cite{DorRodTer84-85}.

The presence of strong magnetic fields can also stimulate the
proton decay
\begin{equation}
p\rightarrow n+e^{+} +\nu_e.
\end{equation}
The decay rate of this process in the strong magnetic field and
the corresponding astrophysical consequences have been discussed
in \cite{BanRub91} (see also \cite{GraRub01}). The proton decay
induced by different configurations of strong electromagnetic
fields has been also considered in \cite{Gaz65,Zar65,Stu83}.

The present paper is devoted to a detailed study of the inverse
beta-decay of neutron in a magnetic field
\begin{equation}\label{proc}
\nu _{e} + n \rightarrow p + e ^{-}.
\end{equation}
The process $\nu  n \rightarrow p e$ in a magnetic field has been
discussed previously by several authors. The contribution of this
process to the conditions for beta-equilibrium in the presence of
magnetic fields has been considered in \cite{CheSchTru93}. The
dependence of the cross section on the magnetic field has been
also discussed \cite{BisKog97} in the context of the pulsar kick
in the case when the asymmetric magnetic field arises just after
the star collapse.

A reasonable interest to the inverse beta-decay of neutron in
magnetic fields has been stimulated by a believe that it can be
relevant for the neutrino opacity in the proto-neutron star stage
after supernova collapse. The first detailed evaluation of the
magnetic field effect on the neutrino opacity can be found in
\cite {Rou97}. In  \cite {Rou97}, as well as in \cite{LaiQia98},
the calculations for the cross section have been performed under
the assumption that the magnetic field gives contribution to the
phase space integrals only, whereas the process matrix element
have been considered unaffected by the magnetic field.

The first attempt to calculate modification of the neutrino sphere
in pulsar due to the asymmetry in the $\nu n \rightarrow pe$ cross
section accounting for the magnetic field modifications of the
matrix element has been undertaken in \cite{Goy98}. However, in
this paper, as well as in \cite{GvoOgn99}, the transition to the
electron lowest Landau level has been discussed. In the paper
\cite{ArrLai99} the angular asymmetry of the cross section has
been calculated only to the first order in the magnetic field.

An important effect of anisotropy in the cross section of the
inverse beta-decay has been recently considered in a series of
papers \cite{BhaPal00,BhaPal00_1,BhaPal02_1} where the process
$\nu n \rightarrow pe$ has been studied in presence of a
background magnetic field and the initial neutron polarization has
been also accounted for.  However, some of the final results of
refs.\cite{BhaPal00,BhaPal00_1} for the cross section do not
coincide with corresponding results of ref. \cite{BhaPal02_1}.
There are also some discrepancies in the figures of
ref.\cite{BhaPal02_1}. It was also claimed in
\cite{BhaPal00,BhaPal02_1} that the developed approach is valid if
the strength of magnetic field is much smaller than
$B_{p}={m^{2}_{p} \over e}\approx 1.5\times 10^{20} G$ and,
therefore, the proton momentum quantization and proton recoil
motion can be neglected. However, this is not an exact estimation
for the upper limit of the magnetic field for which the performed
in \cite{BhaPal02_1} calculations can be applied. The magnetic
field limit depends on the neutrino energy and, as we show below
in the present paper, the expressions for the cross section of the
inverse beta-decay obtained in \cite{BhaPal00,BhaPal02_1} are
valid if the strength of the magnetic field is much smaller than
the proton critical field $ B^{\prime} _{cr}\sim 10^{18} G$ (for
the range of the neutrino energies $\varkappa\sim 10 MeV$). Note
that if the strength of the background field is of the order or
exceeds the critical value $B^{\prime} _{cr}$ then in the
calculation of the cross section one must certainly account for
the influence of the magnetic field on the proton and consider the
Landau quantization of the proton momentum. Moreover, for the
electrically neutral neutrino-proton-electron matter in
beta-equilibrium the magnetic effects on protons are as important
as those on electrons \cite{LaiSha91}. Thus, in such matter
because of charge neutrality the proton critical field may be
forced to reduce to the level of much smaller electron critical
field $B_{cr}$.

The present paper is devoted to a detailed evaluation of the
inverse beta-decay of polarized neutron cross section in a
magnetic field. For both of the charged particles ($e$ and $p$)
wave functions we use the exact solutions of the Dirac equation in
the presence of a magnetic field so that we also exactly account
for the magnetic field influence on the proton. The incoming
neutrino is supposed to be relativistic and effects of neutrino
non-zero mass are neglected. We do not set any special limit on
the neutrino energies, however it is supposed that the
four-fermion weak interaction theory is relevant in our case. For
astrophysical applications, and for supernovas in particular, it
is of interest to consider the neutrino energies in the range of
$\varkappa \sim 1-30 MeV$.

 In our consideration we account for the proton momentum
 quantization in the magnetic field and for the proton recoil motion
so that we develop here the relativistic theory of the inverse
beta-decay. We also suppose that the $Z$ and $W$ bosons are not
affected by the magnetic field. The contribution  of nucleons
anomalous magnetic moments in strong magnetic fields is discussed.
The former effect can be easily incorporated into our calculations
by the corresponding shift of the nucleons masses (see also
\cite{LeiPer98,ArrLai99,GraRub01,MamGoud04}). We also show that in
the case of very strong magnetic fields the process, due to the
anomalous magnetic moments, can be forbidden.

\section{Cross section of inverse beta-decay}

We start with the well known four-fermion Lagrangian,
\begin{equation}\label{Lag}
    {\cal L}=\frac{G}{\sqrt{2}}\left[\overline\psi_p\gamma_\mu(1+\alpha\gamma_5)
 \psi_n\right]\left[\overline\psi_e\gamma^\mu(1+\gamma_5)\psi_\nu\right],
\end{equation}
where $G=G_{F}\cos\theta_c$, $\theta_c$ is the Cabibbo angle, and
$\alpha=1.26$ is the ratio of the axial and vector constant. The
total cross section of the process can be written as
\begin{equation}\label{sigma}
  \sigma=\frac{L^3}{T}\sum_{phase\ space}
  |M|^2,
\end{equation}
where summation is performed over the pase space of the final
particles. The matrix element of the process is given by
\begin{equation}\label{M}
  M=\frac{G}{\sqrt{2}}\int\left[\overline\psi_p\gamma_\mu
  (1+\alpha\gamma_5)
 \psi_n\right]\left[\overline\psi_e\gamma^\mu
 (1+\gamma_5)\psi_\nu\right]dxdydzdt.
\end{equation}
We account for the influence of the background magnetic field on
the matrix element (\ref{M}). The corresponding calculations are
performed by using the exact solutions of the Dirac equation in
the magnetic field for the relativistic electron and proton.
Without loss of generality, a constant magnetic field $\vec B$ is
taken along the $z$-direction. We use the notations of our
previous study \cite{Stu89} of the beta-decay of the polarized
neutron in a magnetic field with the proton recoil effects have
been accounted for. We also choose the longitudinal in respect to
the magnetic field vector $\vec B$ component of the polarization
tensor
\begin{equation}\label{mu_3}
\mu_{3}=m\sigma_{3} +\rho_{2} \Big[\vec \sigma \times \vec
P\Big]_{z},\ \vec P=\vec p+e\vec A, \ \ \
\rho_2=\begin{pmatrix}{0} & {-iI}
\\ {iI} & {0} \end{pmatrix},
\end{equation}
for classifying of the charged particles spin states. Here
$\sigma$ are the Pauli matrixes, $\vec p$ is the momentum of the
particle, $\vec A$ is the vector potential of the magnetic field,
and $0,\ I$ are the $2 \times 2$-matrixes. The detailed discussion
on the derivation of the solution of the Dirac equation in a
magnetic field, and also on different spin operators that can be
used for charged particles in this case, can be found in
\cite{SokTer68}.  The electron wave function
$\psi_e(m,n,s,p_0,p_2,p_3)$ can be written
 in the form
\begin{equation}\label{Ksi_e}
 \psi_e=\frac{1}{L}
 \left(
  \begin{array}{c}
    C_1 U_{n-1}(\eta) \\
    i C_2 U_n(\eta) \\
    C_3 U_{n-1}(\eta) \\
    iC_4 U_n(\eta)
  \end{array}
 \right)e^{-i(p_0 t-p_2 y-p_3 z)},
  \eta=x\sqrt\gamma+\frac{p_2}{\sqrt\gamma},\quad \gamma=eB,
\end{equation}
where $U_n(\eta)$ are Hermite functions of order $n$, $e$ is the
absolute value of the electron charge, $p_0, p_2$ and $p_3$ are
the electron energy and momentum components, respectively. The
energy spectrum
\begin{equation}\label{Enrg_e}
  p_0=\sqrt{m^2+2\gamma n+p_3^2},
\end{equation}
depends on the discreet number $n=0,1,2,...$ denoting the Landau
levels ($m$ is the electron mass). If one uses the spin operator
(\ref{mu_3}) then the spin coefficients $C_i$ are
\begin{equation}\label{C_e}
  C_{1,3}=\frac{1}{2}\sqrt{
  1+s\frac{m\mathstrut}{\tilde p_\bot}}\
  \sqrt{1\pm s\frac{\tilde p_\bot}{p_0}},
  \quad
  C_{2,4}=\mp\frac{1}{2}s\sqrt{1-s\frac{m\mathstrut}{\tilde p_\bot}}\
  \sqrt{1\mp s\frac{\tilde p_\bot}{p_0}},
\end{equation}
and $\tilde p_\bot=\sqrt{m^2+2\gamma n}$. The spin number can have
only the values $\pm1$, $s=+1$ when the electron spin is directed
along the magnetic field $\vec B$, and $s=-1$ in the opposite
case. The electrons on all Landau levels with $n\geq1$ can have
two different spin polarizations. However, in the lowest Landau
state ($n=0$) the electron spin can have the only orientation
given by $s=-1$, so that the electrons moving along the direction
of the magnetic field are left-handed polarized, whereas the
electrons moving in the opposite direction are right-handed
polarized.

The proton wave function $\psi_p(m',n',s',p_0',p_2',p_3')$ can be
expressed in a similar form
\begin{equation}\label{Ksi_p}
 \psi_p=\frac{1}{L}
 \left(
  \begin{array}{c}
    C_1' U_{n'}(\eta') \\
    -i C_2' U_{n'-1}(\eta') \\
    C_3' U_{n'}(\eta') \\
    -i C_4' U_{n'-1}(\eta')
  \end{array}
 \right)e^{-i(p_0't-p_2'y-p_3'z)}.
\end{equation}
The dashed quantities correspond to the proton mass, number of the
Landau state, energy and momentum components. Note that the
positions of $n'$ and $n'-1$ are interchanged compared to the
positions of $n$ and $n-1$ in the electron case. Again the proton
spin values are $s'=\pm 1$, however now at the lowest Landau level
the spin orientation is along the magnetic field $\vec B$.

The initial neutron and neutrino are supposed to be not affected
by the magnetic field, and we use the plane waves for their wave
functions. The polarized neutron wave function can be chosen in
the form
\begin{equation}\label{Ksi_n}
  \psi_{n}=\frac{1}{2L^{3/2}}
 \left(
 \begin{array}{c}
  N_1 \\
  N_2 \\
  N_3 \\
  N_4
 \end{array}
 \right)e^{-i(p_0^n t-\vec p_n\vec r)},
\end{equation}
where the neutron spin coefficients are
\begin{align}\label{Coef-N}
 N_{1,3}&=s_n\sqrt{1\pm \frac{m_n}{p_0^n}}\cdot
 \sqrt{1\pm s_n\cos\theta_n}\cdot e^{\mp i\varphi_n/2},\notag\\
 N_{2,4}&=\sqrt{1\mp \frac{m_n}{p_0^n}}\cdot\sqrt{1\mp s_n\cos\theta_n}
 \cdot e^{\pm i\varphi_n/2}.
\end{align}
Here $m_n,p_0^n$ and $\vec p_n$ are the neutron mass, energy,
momentum, and $\theta_n$, $\varphi_n$ are the polar and azimuthal
neutron momentum angles. The neutron spin value $s_n$ ($s_n=\pm1$)
classifies the neutron states in respect to the spin projection to
$z$-direction ($s_n=1$ corresponds to the spin orientation
parallel to the magnetic field $\vec B$). We perform our
calculations in the rest frame of the neutron, so that we shall
take below $p^{n}_{0}=m_{n}$ and $N_3=N_4=0$.

 The neutrino wave function is
\begin{equation}\label{Ksi_nu}
  \psi_{\nu}=\frac{1}{2L^{3/2}}
 \left(
 \begin{array}{c}
   f_1 \\
   f_2 \\
  -f_1 \\
  -f_2
 \end{array}
 \right)e^{-i(\varkappa t-\vec \varkappa\vec r)},
\end{equation}
where
\begin{equation}\label{f_nu}
  f_1=-e^{-i\varphi_\nu} \sqrt{1-\cos\theta_\nu},
  \quad
  f_2=\sqrt{1+\cos\theta_\nu},
\end{equation}
and $\varkappa,\vec\varkappa$ are the neutrino energy and
momentum, respectively. We neglect effects of the neutrino mass so
that $\varkappa=|\vec\varkappa|$. The neutrino polar and azimuthal
angles are denoted as $\varphi_\nu$ and $\theta_\nu$.

Putting these wave functions to the matrix element of the process
(\ref{M}), we can perform the integrations over time $t$ and
spacial coordinates $y$ and $z$ and obtain three
$\delta$-functions
\begin{multline}\label{Delt0}
  \int e^{-it(m_n +\varkappa -p_0-p_0')+iy(\varkappa_2-p_2-p_2')+
  iz(\varkappa_3-p_3-p_3')}dtdydz=\\
  =(2\pi)^3\delta(p_0'+p_0-m_n-\varkappa)
  \delta(p_2'+p_2-\varkappa_2)\delta(p_3'+p_3-\varkappa_3).
\end{multline}

To integrate over the coordinate $x$ in the matrix element we use
the properties of the Hermite functions (see
\cite{SokTer68,GraRyz80}) and the result
\begin{gather}
  \int\limits_\infty^\infty U_n(\eta)U_{n'}(\eta')e^{-i\varkappa_1
  x}dx=I_{n',n}(\rho)e^{i\mu+i(n-n')\lambda},\label{Prop}\\
  \mu=\frac{\varkappa_1 (p_2+p_2')}{2\gamma},\quad
  \lambda=\arctan\frac{\varkappa_1}{p_2'+p_2},\quad
  \rho=\frac{\varkappa_1^2+(p_2'+p_2)^2}{2\gamma}.\label{rho}
\end{gather}
The Lagguere function $I_{n',n}(\rho)$ is connected with the
Lagguere polynomials $Q^l_n(\rho)$:
\begin{equation}\label{Lag_f}
  I_{n',n}(\rho)=\frac{1}{\sqrt{n'!n!}}e^{-\frac{\rho}{2}}
  \rho^\frac{n'-n}{2}Q_n^{n'-n}(\rho), \quad
  Q_n^l(\rho)=e^\rho\rho^{-l}\frac{d^n}{d\rho^n}
  \left(\rho^{n+l}e^{-\rho}\right).
\end{equation}

Finally for the matrix element of the inverse beta-decay of the
neutron we get
\begin{multline}\label{M_fnl}
  M=i\sqrt{2}(2\pi)^3G\, e^{i\mu+i(n-n')\lambda}
  \Big[\Big\{2(\alpha C_1'-C_3')
  f_1N_2-(\alpha+1)(C_1'-C_3')f_2N_1\Big\}\times\\
  \times(C_2-C_4)I_{n',n}(\rho)-\Big\{2(\alpha C_2'-C_4')f_2N_1-
  (\alpha+1)(C_2'-C_4')f_1N_2\Big\}\times\\
  \times(C_1-C_3)I_{n'-1,n-1}(\rho)+i(\alpha-1)(C_1-C_3)
  (C_1'+C_3')f_1N_1I_{n',n-1}(\rho)e^{-i\lambda}+\\
  +i(\alpha-1)(C_2-C_4)(C_2'+C_4')f_2N_2I_{n'-1,n}(\rho)
  e^{i\lambda}\Big]\\
\times\delta(p_0'+p_0-m_n-\varkappa)
  \delta(p_2'+p_2-\varkappa_2)\delta(p_3'+p_3-\varkappa_3),
  \end{multline}
(here we do not include the overall term ${1 \over {4L^{5}}}$
which shall be accounted for below). Note that this expression for
the matrix element follows from the relativistic matrix element of
the direct beta-decay in the magnetic field calculated in
\cite{Stu89}.

Using the usual rules like
\begin{gather}\label{Delt2}
  |\delta(p_0'+p_0-m_n-\varkappa)|^2=
  \frac{T}{2\pi}\delta(p_0'+p_0-m_n-\varkappa),\\
  |\delta(p_2'+p_2-p_2^n-\varkappa_2)|^2=
  \frac{L}{2\pi}\delta(p_2'+p_2-p_2^n-\varkappa_2),\\
  |\delta(p_3'+p_3-p_3^n-\varkappa_3)|^2=
  \frac{L}{2\pi}\delta(p_3'+p_3-p_3^n-\varkappa_3),
\end{gather}
where $T$ and $L$ are the quantization large time and regions in
the $y$ and $z$ directions, we get the following relation for the
squared norm of the matrix element
\begin{multline}\label{MM}
|M|^2=(2\pi)^{3}TL^{2}|\tilde M|^2\delta(p_0'+p_0-m_n-\varkappa)
\\
\times\delta(p_2'+p_2-\varkappa_2)\delta(p_3'+p_3-\varkappa_3),
\end{multline}
where
\begin{multline}\label{mm}
|\tilde M|^2=2G^2\Big[ (\alpha-1)^2f_1^2 N_1^2
  (C_1-C_3)^2 (C_1'+C_3')^2 {I}_{n',n-1}^2(\rho)+\\
  +(C_2-C_4)^2\Big\{4f_1^2 N_2^2 (\alpha C_1'-C_3')^2
  +(\alpha+1)^2 f_2^2 N_1^2(C_1'-C_3')^2\Big\}I_{n',n}^2(\rho)+\\
  +(\alpha-1)^2 f_2^2 N_2^2(C_2-C_4)^2(C_2'+C_4')^2
  I_{n'-1,n}^2(\rho)+(C_1-C_3)^2\times\\
  \times\Big\{4f_2^2 N_1^2(\alpha
  C_2'-C_4')^2+(\alpha+1)^2 f_1^2 N_2^2 (C_2'-C_4')^2\Big\}
  I_{n'-1,n-1}^2(\rho)+\\
  +4(\alpha+1)(C_1-C_3)(C_2-C_4)\Big\{f_2^2 N_1^2
  (C_1'-C_3')(\alpha C_2'-C_4')+\\
  +f_1^2 N_2^2 (C_2'-C_4')
  (\alpha C_1'-C_3')\Big\}I_{n',n}(\rho)I_{n'-1,n-1}(\rho)\Big].
\end{multline}

Let us now return back to the general expression (\ref{sigma}) for
the cross section of the process and perform the integration and
summation over the phase space of the final particles. The phase
space factor for the electron and proton in the presence of a
magnetic field is
\begin{equation}\label{ph_f}
 \sum_{phase \ space}=\int {L\over 2\pi}dp_{2}{L\over 2\pi}dp_{3}
 {L\over 2\pi}dp'_{2}{L\over 2\pi}dp'_{3}
 \sum_{n=0, \\ n'=0}\sum_{s=\pm 1,s'=\pm 1}g_{n}g_{n'},
\end{equation}
where $g_0=1$, and $g_k=2$ for $k\geq 1$ are the degeneracies of
the Landau energy levels for the electron and proton. The
integrations over the proton momentum component $p'_2$ and the
electron momentum component $p_3$ are performed by using of the
two delta-functions $\delta(p_2'+p_2-\varkappa_2)$ and
$\delta(p_3'+p_3-\varkappa_3)$, respectively. After these
integrations we get the laws of conservation for the two momentum
components, $p_3=\varkappa_3-p_3',\ p'_2=\varkappa_2-p_2$.

The integration over the electron momentum component $p_2$ is
performed by taking in to account the specific for the motion in a
magnetic field degeneracy of the electron energy. The
corresponding phase space factor is
\begin{equation}\label{int_p2}
  \int\limits_\infty^\infty dp_2\rightarrow \frac{2\pi}{L}\sum_{p_2}
  \rightarrow eBL.
\end{equation}

Finally we obtain the cross section of the inverse beta-decay of
the polarized neutron in a magnetic field with the proton recoil
motion effect being accounted for,
\begin{equation}\label{fin}
  \sigma=\frac{eB}{32\pi}
  \sum_{s, s'}\sum_{n,n'}\int\limits_\infty^\infty
  {|\tilde M|}^2
  \delta_0(p_0'+p_0-m_n-\varkappa)\big|_{p_3=\varkappa_3 - p'_3}dp_3',
\end{equation}
where $|\tilde M|^2$ is given by (\ref{mm}) with $p_3$ being
substituted by $\varkappa_3 - p'_3$ because at this stage of
calculations we have already perform the integration over the
component of the electron momentum $p_3$ with use of the
corresponding $\delta$-function.

The remaining integration over the component $p'_3$ of the proton
momentum is performed with use of the $\delta_{0}\big(\varphi
(p'_3)\big)$-function. The argument $\varphi (p'_3)$, being
equated with zero, gives the law of energy conservation for the
particles in the process,
\begin{equation}\label{energy}
m_{n}+\varkappa={\sqrt{m^2+2\gamma
n+(\varkappa_{3}-p'_{3})^2}}+{\sqrt{m'^2+2\gamma n'+{p'_{3}}^2}}.
\end{equation}
Thus, the argument of the $\delta_0$-function in (\ref{fin}) is a
complicated function of $p'_3$. That is why in order to perform
the integration over $p'_3$ we have to use the relation
\begin{equation}\label{delta}
  \delta\Big(\varphi(p'_{3})\Big)=\sum_i
  \frac{\delta(p'_{3}-{p'_{3}}^{(i)})}{|\varphi'({p'_{3}}^{(i)})|},
\end{equation}
where
\begin{equation}\label{derivative}
\varphi'({p'}_{3}^{(i)})\equiv \frac{d \varphi ({p'}_{3})}{d
p'_{3}}\Big|_{{p'}_{3}= {p'}_{3}^{(i)}},
\end{equation}
and ${p'_{3}}^{(i)}$ are the simple roots of the equation
\begin{equation}\label{phi}
  \varphi({p'_{3}}^{(i)})=0.
\end{equation}
For the argument of the $\delta_{0}\big(\varphi
(p'_3)\big)$-function in (\ref{fin}) we get
\begin{equation}\label{fi}
  \varphi(p_3')=\sqrt{\tilde{p}_\bot'^2+p_3'^2}+\sqrt{\tilde{p}_\bot^2+(
  \varkappa_3-p_3')^2}-m_n-\varkappa,
\end{equation}
then the derivative is
\begin{equation}
\varphi'(p_3')=\frac{p_3'}{\sqrt{\tilde{p}_\bot'^2+p_3'^2}}-
  \frac{\varkappa_3-p_3'}{\sqrt{\tilde{p}_\bot^2+(\varkappa_3-p_3')^2}},
\end{equation}
where
\begin{equation}\label{p_perp}
\tilde p'_\bot=\sqrt{m'^2+2\gamma n'},
  \tilde p_\bot=\sqrt{m^2+2\gamma n}.
\end{equation}
 There are the two roots of the equation (\ref{phi}),
\begin{multline}\label{roots}
  p_3'^{(1,2)}=\frac{1}{2\Big[(m_n+\varkappa)^2-\varkappa_3^2\Big]}
  \Bigg\{\varkappa_3\Big[(m_n+\varkappa)^2+\tilde{p}_\bot'^2-
  \tilde{p}_\bot^2-\varkappa_3^2\Big\} \\
  \pm (m_n+\varkappa)\sqrt{\Big[(m_n+\varkappa)^2-\tilde{p}_\bot'^2-
  \tilde{p}_\bot^2-\varkappa_3^2 \Big]^2-
  4\tilde{p}_\bot'^2\tilde{p}_\bot^2} \Bigg\}.
\end{multline}

Finally we obtain the cross section of the inverse beta-decay of
the polarized neutron in a magnetic field, with the effects of the
Landau quantization of the proton momentum and of the proton
recoil motion being accounted for exactly,
\begin{equation}\label{cross}
  \sigma=\frac{eB}{32\pi}
  \sum_{s, s'}\sum_{n,n'}\sum_{i=1,2}
  {{{|\tilde {M}^{(i)}|}^{2}} \over {
  \Big|{{p_{3}^{(i)} \over p_{0}^{(i)}}-{p_{3}'^{(i)}
  \over p_{0}'^{(i)}}}\Big|}},
\end{equation}
where one of the sums is performed over the roots $p_3'^{(i)}$ of
equation (\ref{phi}) given by (\ref{roots}), and
\begin{gather}
  p_3^{(i)}=\varkappa_3-p_3'^{(i)},
  \quad p_0'^{(i)}=\sqrt{\tilde{p}_\bot'^2+p_3'^{(i)2}},\quad
  p_0^{(i)}=\sqrt{\tilde{p}_\bot^2+(\varkappa_3-p_3'^{(i)})^2}.
   \label{En-Im}
\end{gather}
The squared matrix element ${|\tilde {M}^{(i)}|}^2$ is given by
eq. (\ref{mm}) where the substitution ${p_{3}'}\rightarrow
{p'_{3}}^{(i)}$ must be done,
\begin{equation}
{|\tilde {M}^{(i)}|}^2={\tilde{|M|}^2}_
{\big|_{{p_{3}'}={p'_{3}}^{(i)}}}.
\end{equation}

Now let us consider eq. (\ref{energy}) in detail that gives the
energy conservation law by accounting for the presence of a
magnetic field. Due to the particular properties of the energy
spectra of the electron and proton in a magnetic field, we can
introduce two critical values of the magnetic field strength.
First let us determine the critical electron magnetic field,
$B_{cr}$, from the condition that in the external field $B\geq
B_{cr}$ the electron can occupy only the lowest Landau level with
the number $n=0$. From (\ref{energy}) we get that for a fixed
maximum neutrino energy $\varkappa_{max}$ and for a fixed strength
of the magnetic field, the maximum number of the available
electron Landau level is

\begin{equation}\label{n_max}
n_{max}=\text{int} \Big[{{(\Delta+\varkappa_{max})^{2}-m^2} \over
{2eB}}\Big],
\end{equation}
where $\Delta=m_n-m'$ is the difference in masses of the neutron
and proton. From the condition $n_{max}<1$ (it means that the
electron can occupy only the lowest Landau level with $n=0$) we
get
\begin{equation}\label{B_cr}
  B_{cr}={{(\Delta+\varkappa_{max})^{2}-m^2} \over {2e}}.
\end{equation}
Thus, $B_{cr}$ depends on the maximum available neutrino energy.
For example, for different neutrino energies we have the following
values of the electron critical magnetic field:
\begin{equation}\label{}
B_{cr}\approx 8.3 \times 10^{16} G, \ \ \varkappa_{max}=30 \ MeV,
\end{equation}
\begin{equation}\label{}
B_{cr}\approx 1.1 \times 10^{16} G, \ \ \varkappa_{max}=10 \ MeV,
\end{equation}
\begin{equation}\label{}
B_{cr}\approx 1.2 \times 10^{14} G, \ \ \varkappa_{max} \ll \ m.
\end{equation}

The critical proton magnetic field, $B'_{cr}$, was determined from
the condition that in the external field $B\geq B'_{cr}$ the
proton can occupy only the lowest Landau level with the number
$n'=0$. Again, from (\ref{energy}) we get that for the fixed
maximum neutrino energy $\varkappa_{max}$ and for the fixed
strength of the magnetic field the maximum number of the available
proton Landau level is

The critical proton magnetic field, $B'_{cr}$, was determined from
the condition that in the external field $B\geq B'_{cr}$ the
proton can occupy only the lowest Landau level with the number
$n'=0$. Again, from (\ref{energy}) we get that for a fixed maximum
neutrino energy $\varkappa_{max}$ and for a fixed strength of the
magnetic field, the maximum number of the available proton Landau
level is

\begin{equation}\label{n'_max}
n'_{max}=\text{int} \Big[{(\varkappa_{max} +m_n-m)^{2}-m'^2 \over
{2eB}}\Big].
\end{equation}
The proton can occupy only the Landau level with $n'=0$ if the
magnetic field strength exceeds the proton critical field
\begin{equation}\label{B'_cr}
B'_{cr}={(\varkappa_{max} +m_n-m)^{2}-m'^2 \over {2e}}.
\end{equation}
For different energies of the incoming neutrino we get
\begin{equation}\label{}
B'_{cr}\approx 5 \times 10^{18} G, \ \  \varkappa_{max}=30 \ MeV
\end{equation}
\begin{equation}\label{}
B'_{cr}\approx 1.7 \times 10^{18} G, \ \ \varkappa_{max}=10 \ MeV
\end{equation}
\begin{equation}\label{}
B'_{cr}\approx 1.3 \times 10^{17} G, \ \ \varkappa_{max} \ll \ m.
\end{equation}

In Fig.1 we plot the values of the critical fields $B_{cr}$
(dashed line) and $B'_{cr}$ (solid line) as functions of the
initial neutrino energy $\varkappa$.

From the above we conclude that there are the three ranges of the
magnetic field strength which we call: 1) the weak field ($B\leq
B_{cr}$), 2) the strong field ($B_{cr}< B< B'_{cr}$), and 3) the
super-strong field ($B'_{cr}\leq B$). For the most of the weak
field range ($B\ll B_{cr}$) the electron $n$ and proton $n'$
Landau numbers can have very large values. Inside the strong field
range ($B_{cr}< B\ll B'_{cr}$) only the proton number $n'$ can
have very large values, whereas the electron number is always
zero. In the super-strong fields the both Landau numbers are zero,
$n=n'=0$.

The electron and proton spin properties are very different for
each of the three ranges of the magnetic field strengths. In weak
fields the electron and proton can have the two spin
polarizations, in the strong (and the super-strong) fields the
electron is always polarized against the direction of the magnetic
field, and in the super-strong fields the electron and proton spin
polarizations are opposite. In this concern it is reasonable to
expect that the expressions for the differential cross sections
for the inverse beta-decay in these three ranges of magnetic
fields are very different. This, in particular, have to be
reflected in the dependence of the cross section on the
polarizations of the particles and also on asymmetries in respect
to the neutrino angle $\theta$. Also it is reasonable to expect
that the expression for the cross section calculated in the
presence of the strong field cannot be applied to the case of the
super-strong magnetic field.

\section{Cross section in super-strong, strong and weak magnetic
fields}

We consider the cross section of the inverse beta-decay of the
polarized neutron, accounting also for the proton recoil motion,
in the three ranges of the background field: $ 1) B\leq B_{cr},\
2)B_{cr}< B< B'_{cr},\ 3)B'_{cr}\leq B$. Here we should like to
emphasize that in the previous section we have derived the general
expression (\ref{cross}) for the cross section, accounting for the
proton recoil motion, that can be used for arbitrary magnetic
fields.  As we have already discussed in Section 1, the energy
spectra of the electron and also the proton are quantized into the
Landau levels in the presence of a magnetic field. These specific
properties of the energy spectra of the charged particles set the
three rather different methods of the further analytical
calculations of the cross section using the general expression
(\ref{cross}).

\subsection{Cross section in super-strong magnetic field}
Let us start with consideration of the super-strong magnetic field
$B\geq B'_{cr}$. Obviously, in this case the calculations are
reasonably simplified because, as it has been already discussed
above, the both numbers of the Landau levels for the electron and
proton are zero. The Laguerre functions (\ref{Lag_f}) for $n=n'=0$
are
\begin{equation}\label{Lag_00}
I_{0,0}(\rho)=e^{-{\rho \over 2}},
\end{equation}
where the argument is
\begin{equation}\label{rho_00}
  \rho=\frac{\varkappa_1^2+\varkappa_2^2}{2\gamma}=\frac{\varkappa_\bot^2}{2\gamma},
  \quad
    \varkappa_\perp=\sqrt{\varkappa^2-\varkappa_3^2}=\varkappa\sin\theta,
\end{equation}
and the squared matrix element $|\tilde M|^2$ in eq. (\ref{mm}) is
reduced to
\begin{multline}\label{mm_00}
  |\tilde M_{n{=}n'{=}0}|^2=2G^2(C_2-C_4)^2\Big\{4f_1^2 N_2^2
  \alpha C_1'-C_3')^2+\\
  +(\alpha+1)^2 f_2^2 N_1^2 (C_1'-C_3')^2\Big\}e^{-\rho}.
\end{multline}
Putting back in (\ref{cross}), we obtain the cross section of the
process in the presence of the super-strong magnetic field
\begin{multline}\label{cross_s_00}
 \sigma_{n{=}n'{=}0}=\frac{eBG^2}{8\pi} e^{-\frac{\varkappa_\bot^2}
 {2\gamma}}\sum_{i=1,2}{\frac
 {\Big(1+\frac{p_{3}^{(i)}}{p_{0}^{(i)}}\Big)}
  {\Big|{p_{3}^{(i)} \over p_{0}^{(i)}}-{p_{3}'^{(i)}
 \over p_{0}'^{(i)}}\Big|}}
 \Big\{a^{(i)} +b^{(i)}\cos\theta
 +s_{n}(b^{(i)}+a^{(i)}\cos\theta)\Big\},
\end{multline}
where
\begin{gather}
  a^{(i)}=3+2\alpha+3\alpha^2-2(1-\alpha^2)\frac{m'}{p_{0}'^{(i)}}-
  (1+6\alpha+\alpha^2)\frac{p_{3}'^{(i)}}{p_{0}'^{(i)}},
  \notag \\ \label{a_b}
  b^{(i)}=-1+2\alpha -\alpha^2+2(1-\alpha^2)\frac{m'}{p_{0}'^{(i)}}-
  (1-\alpha)^2\frac{p_{3}'^{(i)}}{p_{0}'^{(i)}}.
\end{gather}
The effect of the proton motion, which appears in this case
exceptional due to the proton recoil in z-direction, is accounted
exactly in eqs. (\ref{cross_s_00}) and (\ref{a_b}). It is worth to
be  mentioned that the derived expression for the cross section in
the super-strong magnetic field $B\geq B'_{cr}$ can be applied for
neutrinos with arbitrary (also ultra-high) energies (note that
following to eq. (\ref{B'_cr}) the value of $B'_{cr}$ is
increasing with the neutrino energy increase).
%\footnote{Note that following to eq. (\ref{B'_cr}) the value of
%$B'_{cr}$ is increasing with the neutrino energy increase.}.

If we neglect the proton momentum parallel or antiparallel to the
magnetic field, we get
\begin{equation}\label{cross_s_00'}
  {\sigma_{{n{=}n'{=}0}}}_{{\big |}_{ p'_{0}=m'}}=\frac{eBG^2}{4\pi}
e^{-\frac{\varkappa_\bot^2}
 {2\gamma}}
 \{a+b\cos\theta+s_{n}(b+
  a\cos\theta)\}\frac{\Delta+\varkappa}{\sqrt{(\Delta+\varkappa)^2-m^2}},
\end{equation}
where
\begin{equation}\label{a0-b0}
  a=1+2\alpha+5\alpha^2,
  \quad
  b=1+2\alpha-3\alpha^2.
\end{equation}
For $\alpha = 1.26$ one can get $a = 11.5$ and $ b= -1.24$. Note
that the same coefficients $a$ and $b$ determine the neutrino
asymmetry in the probability of the direct neutron beta-decay in
the super-strong magnetic field \cite{Stu89,Stu88}.

In the case of moderate neutrino energies $\varkappa^{2}\ll eB$
(the last inequality is valid in the super-strong magnetic field
$B\geq B'_{cr}$ for the range of the neutrino energies $\varkappa
\leq 30 MeV$) the exponential term in (\ref{cross_s_00}) must be
substituted for unit. The coefficients $a_0$ and $b_0$ for the
cross section of the inverse beta-decay of the neutron in the case
when $n=n'=0$ and for the neutrino energies  $\varkappa^{2}\ll eB$
have been also obtained in \cite{Goy98}.

As it follows from the used neutron wave function, eqs.
(\ref{Ksi_n}) and (\ref{Coef-N}), the neutron spin quantization
axis is parallel to the magnetic field vector $\vec B$. Therefore,
the derived above expressions for the cross section, which
contains the value $s_n=\pm 1$, describe the neutrino interaction
with neutrons totally polarized along ($s_n=+1$) or against
($s_n=-1$) the magnetic field. In the case of non-polarized
neutrons we have to overage the cross section over the neutron
spin
\begin{equation}\label{us}
{\sigma_{unpol.}}= \frac{1}{2} {\sum_{s_n=\pm 1}} \sigma (s_n).
\end{equation}
We also can use the obtained expressions for the cross section in
analysis of the neutrino interaction with partially polarized
neutron matter when the numbers of neutrons (per unit volume) with
the two different spin polarizations are $N_{+}$ and $N_{-}$,
respectively. The partially polarized neutron matter can be
characterized by the neutrons polarization $S$ determined as
\begin{equation}\label{S}
  S=\frac{N_{+}-N_{-}}{N_{+}+N_{-}}.
\end{equation}
All the obtained above formulas for the cross section can be used
for the case of partially polarized neutron matter if one
substitutes $s_n$ for $S$.

In Figs.2 (a,b,c), we have plotted, for different neutrino
energies $\varkappa= 30\ MeV, 10\ MeV$ and $\varkappa\ll m$, the
cross section in the magnetic field $B=B'_{cr}$, normalized to the
cross section in the field-free case, as a function of neutrons
polarization $S$ and of $\cos \theta$ ($\theta$ is the angle the
neutrino momentum makes with the magnetic field). It is clearly
seen that the cross section depends on the direction of the
neutrino momentum and the neutrons polarization. The most
considerable increase (by the factors from a few tenth up to
hundreds depending on the initial neutrino energy) of the cross
section in $B=B'_{cr}$ appears in the two cases of: 1) nearly
total neutrons polarization parallel to the magnetic field
($S\approx 1$) and neutrino propagation parallel to the magnetic
field $\cos \theta \approx 1$), and 2) nearly total neutrons
polarization antiparallel to the magnetic field ($S\approx -1$)
and neutrino propagation antiparallel to the magnetic field ($\cos
\theta \approx -1$).

On the opposite, the cross section (\ref{cross_s_00'}) vanishes to
zero for the cases when the direction of the neutrons total
polarization is antiparallel to the direction of the neutrino
momentum, $S\cos\theta=-1$. For $\cos\theta=+1$ from
(\ref{cross_s_00'}) we get
\begin{equation}\label{}
  {\sigma_{{n{=}n'{=}0}}}_{{\big |}_{ p'_{0}=m',\cos\theta=1}}
  =\frac{eBG^2}{2\pi}
  e^{-\frac{\varkappa_\bot^2}
 {2\gamma}}(1+\alpha^2)(1+S)\frac{\Delta+\varkappa}{\sqrt{(\Delta+\varkappa)^2-m^2}}.
\end{equation}
Therefore, the cross section is zero if $S=-1$. For
$\cos\theta=-1$ from (\ref{cross_s_00'}) we get
\begin{equation}\label{}
  {\sigma_{{n{=}n'{=}0}}}_{{\big |}_{ p'_{0}=m',\cos\theta=-1}}
  =\frac{eBG^2}{\pi}
  e^{-\frac{\varkappa_\bot^2}
 {2\gamma}}2\alpha^2(1-S)\frac{\Delta+\varkappa}{\sqrt{(\Delta+\varkappa)^2-m^2}}.
\end{equation}
Therefore, the cross section is again zero if $S=+1$.

Thus, for these two cases the neutron matter is transparent for
neutrinos. These phenomenon appears due to the Landau quantization
of the momentum and  the spin properties of the charged particles
in the strong and super-strong magnetic fields. In the field
$B\geq B'_{cr}$ the final electron and proton can move only
parallel to the fixed line that is given by the magnetic field
vector. For the the neutrino also moving along this line and the
neutron being at rest, the law of angular momentum conservation
reduces to the law of "spin number conservation". Since the sum of
spin numbers of the initial particles is equal to $\pm2$, whereas
the sum of spin numbers of the final particles is zero in the two
considered cases, the cross sections must vanish. In Fig.3 we
present an illustration of the law of the "spin number
conservation".

The dependence of the cross section on the magnetic field strength
$B\geq B'_{cr}$ for different neutrino energies ($\varkappa = 30\
MeV,\ 10\ MeV$ and $\varkappa \ll m$) and the two directions of
the neutrino momentum ($\cos \theta = \pm 1$) in the case of
unpolarized neutrons ($S=0$) is shown in Fig.4 (a,b,c). In Fig.5
(a,b,c) and Fig.6 (a,b,c), we have plotted the cross sections for
the cases of the totally polarized neutrons ($S=\pm1$). As we have
already discussed above, neutrinos freely escape from the neutron
matter when they move antiparallel to the neutrons polarization,
i.e. $S\cos \theta=-1$.

\subsection{Cross section in strong magnetic field}

In the case of strong magnetic fields $B_{cr}\leq B< B'_{cr}$, the
electron can only occupy the lowest Landau level with $n=0$,
whereas there could be many Landau levels available for the
proton. The maximum number of the proton Landau level is estimated
as
\begin{equation}\label{N'_max}
  n'_{max}=\text{int}\Bigg[\frac{(m_n+\varkappa-m)^2-m'^2}{2eB}\Bigg]\approx
  \text{int}\Bigg[\frac{m'(\Delta+\varkappa-m)}{eB}\Bigg].
\end{equation}
For the squared matrix element of the process we get from
(\ref{mm})
\begin{multline}\label{M_n_0}
  |{\tilde {M}}_{n{=}0}|^2=2G^2(C_2-C_4)^2\Big[
  (\alpha-1)^2 f_2^2 N_2^2(C_2'+C_4')^2{I_{{n'-1,0}}^2}(\rho)+\\
  +\Big\{4f_1^2 N_2^2(\alpha C_1'-C_3')^2+(\alpha+1)^2
  f_2^2 N_1^2(C_1'-C_3')^2\Big\}{I_{{n',0}}^2}(\rho)\Big],
\end{multline}
where the Laguerre functions with $n=0$ are
\begin{equation}\label{I_n_0}
  I_{n',0}(\rho)=\frac{1}{\sqrt{n'!}}x^{\frac{n'}{2}}e^{-\frac{\rho}{2}}.
\end{equation}
Putting the squared matrix element (\ref{M_n_0}) to the general
formula for the cross section, eq. (\ref{cross}), we get the
expression for the cross section,
\begin{multline}\label{cross_strong}
  \sigma_{n{=}0}=\frac{eB G^2}{8\pi}
  \sum_{n'=0}
  ^{n'_{max}}\sum_{i=1,2}{\frac
 {\big(1+\frac{p_{3}^{(i)}}{p_{0}^{(i)}}\big)}
  {\Big|{p_{3}^{(i)} \over p_{0}^{(i)}}-{p_{3}'^{(i)}
 \over p_{0}'^{(i)}}\Big|}}
  \Bigg\{
  \bigg[(1+\alpha)^2\Big(1-\frac{p_3'^{(i)}}{p_0'^{(i)}}
  \Big)
  (1+S)(1+\cos\theta)\\
   +2\Big[1+\alpha^2-(1-\alpha^2)
  \frac{m'}{p_0'^{(i)}}-
  2\alpha\frac{p_3'^{(i)}}{p_0'^{(i)}}
  \Big](1-S)(1-\cos\theta)\bigg]{I_{{n',0}}^2}(\rho) \\
  +(1-\alpha)^2
  \Big(1-\frac{p_3'^{(i)}}{p_0'^{(i)}}\Big)
  (1-S)(1+\cos\theta)(1-\delta _{n',0}){I_{{n'-1,0}}^2}(\rho)
  \Bigg\},
\end{multline}
where $\delta _{n',0}$ is the Kronecker delta ($1-\delta
_{n',0}=0$ for $n'=0$).

Together with the expression (\ref{roots}) for ${p'_3}^{(i)}$,
equation (\ref{cross_strong}) gives the cross section for the
process in the strong magnetic field exactly accounting for the
proton momentum quantization and the proton recoil motion. It
should be noted that in derivation of (\ref{cross_strong}) we have
not use any constraints on the neutrino energy. Thus, eq.
(\ref{cross_strong}) can be used for the case of high energy
neutrino.

Alternatively, if the initial neutrino energy is much less than
the proton mass, $\varkappa\ll m'$, it is possible to get an
approximate analytical expression for the cross section accounting
for the proton recoil motion using the prescription that we have
developed in the study \cite{Stu89} of the proton recoil motion
effect in the beta-decay of the neutron in a magnetic field. The
proton recoil motion can be characterized by the two parameters,
$\alpha' _\parallel = {p'_3 \over m'}$ and $\alpha' _\perp =
{p'_\perp \over m'}$. The maximum values of these parameters are
determined by the initial neutrino energy, and   in the mentioned
above neutrino energy range $\alpha' _\parallel$, $\alpha' _\perp
\ll 1$. Therefore, in order to account for the proton recoil
motion one has to expand in (\ref{fin}), prior to integration over
$p'_3$, the $\delta$-function over the parameter $\alpha' _\perp$,
\begin{gather}\label{delta_exp}
  \delta(m_n+\varkappa-p_0'-p_0)\approx\delta(m_n+\varkappa-p_0'-p_0)+
  \frac{\gamma n'}{\tilde p_3'}\delta'(m_n+\varkappa-p_0'-p_0)+
  \textit{O}\Big({{p'_{\parallel}}^{4} \over {{\tilde {p_3'}}^5}}\Big),
\end{gather}
where $\tilde p'_3=\sqrt{{m'}^2+{p'_3}^2}$. For the case of
magnetic fields $B\ll B'_{cr}$ the maximum number of the proton
Landau level is $n'_{max}> 10$. Thus, it is possible shift the
upper limit $n'_{max}$ in the summation over $n'$ to infinity,
\begin{equation}\label{sum_n'}
\sum _{n'=0}^{n'_{max}}\rightarrow \sum _{n'=0}^{\infty}.
\end{equation}
In addition, if one also performs expansion over the parameter
$\alpha' _\parallel$, then it will be possible to calculate the
sum over the proton Landau number $n'$ (see for details in
\cite{Stu89}) and get the cross section that accounts for the
transversal and longitudinal proton motion in the linear
approximation. The final expression, however, is rather
complicated in this case and we do not present it in this paper.

A reasonable simplification can be achieved if we neglect the
proton motion in the plain orthogonal to the magnetic field vector
and account only for the proton recoil in $z$-direction. In this
case we extract the zeroth-order term in the expansion of the
cross section (\ref{fin}) over the parameter $\alpha' _{\bot}$ and
get,
\begin{multline}\label{cross_n_0_n'_0}
  {\sigma_{n{=}0}}_{{\big |}_{p'_{\bot}=0, p_3'{\neq}0}}=
  \frac{ eB G^2}{4\pi}
  \sum_{i=1,2}{\Big(1+\frac{p_3^{(i)}}{p_0^{(i)}}\Big)\over
  {\Big|{p_{3}^{(i)}
  \over p_{0}^{(i)}}-{p_{3}'^{(i)}
 \over {\tilde {p}}_{3}'^{(i)}}\Big|}}
  \\
  \times \Bigg\{\Big(1+3\alpha^2
  -(1+\alpha)^2\frac{p_3'^{(i)}}{m'}\Big)+
  \Big(1-\alpha^2-(1-\alpha)^2\frac{p_3'^{(i)}}{m'}\Big)\cos\theta\\
  +S\bigg[2\alpha(1-\alpha)
  +\Big(2\alpha(1+\alpha)-4\alpha\frac{p_3'^{(i)}}{m'}
    \Big)\cos\theta\bigg]\Bigg\}.
\end{multline}

Note that the last expression does not reproduce the cross section
$\sigma_{n=n'=0}$ given by (\ref{cross_s_00}) because in eq.
(\ref{cross_n_0_n'_0}), contrary to eq. (\ref{cross_s_00}),
contributions from infinitely many proton Landau levels are
included. Since both spin states $s'=\pm 1$ are not excluded now,
there is only one set of values ($\cos \theta, S$), that determine
the direction of the neutrino momentum ($\cos \theta$) and
polarization of the neutrons ($S$), for which the cross section is
vanish. The final particles total spin number $s+s'$ can be equal
to $0$ or $-2$, whereas for $\cos \theta =\pm 1$ and $S=\pm1$ the
initial particles total spin number can be equal to $0$ or $\pm
2$. Therefore, the violation of the angular momentum conservation
can appear only if $\cos\theta =-1$. The cross section in this
case is
\begin{multline}\label{cross_theta_-1}
  {\sigma_{n{=}0}}_{{\big |}_{p_3'{\neq}0,\ \cos \theta=-1}}=
  \frac{ eB G^2}{\pi}
  \sum_{i=1,2}{\Big(1+\frac{p_3^{(i)}}{p_0^{(i)}}\Big)\over
  {\Big|{p_{3}^{(i)}
  \over p_{0}^{(i)}}-{p_{3}'^{(i)}
 \over {\tilde {p}}_{3}'^{(i)}}\Big|}}
    (1-S)\Big(\alpha^2-\alpha {p'^{(i)} \over m'}\Big),
\end{multline}
and vanishes, as a consequence of the law of "spin number
conservation", if neutrons are polarized in $+z$-direction, i.e.
$S= 1$.

%one considers the case when proton is at the lowest Landau level
%with $n'=0$, so that the proton can move only in $z$-direction.
%This particular case is of interest because it corresponds to
%($n-p-e$) matter in $\beta$-equilibrium in the presence of strong
%magnetic field. If the magnetic field is $B\geq B_{cr}$, so that
%the electrons occupy the lowest Landau level $n=0$, then, as it
%was shown in \cite{LaiSha91}, because of charge neutrality of
%matter, the protons too are forced to occupy the lowest level
%$n'=0$.

If we also neglect the effect of the proton motion in
$z$-direction, then for the cross section in the strong field
$B_{cr}< B\ll B'_{cr}$ we get

\begin{multline} \label{cross_n_0_n'_0_p'_0}
  {\sigma_{n{=}0}}_{{\big |}_{p'_0{=}m'}}=\frac{eB G^2}{2\pi}\bigg\{1+3\alpha^2+
  (1-\alpha^2)\cos\theta+\\
  +2\alpha S\Big[1-\alpha+(1+\alpha)\cos\theta\Big]
  \bigg\}\frac{\Delta+\varkappa}{\sqrt{(\Delta+\varkappa)^2-m^2}}.
\end{multline}
This result for the cross section reproduces the one of ref.
\cite{BhaPal02_1}.

From (\ref{cross_n_0_n'_0_p'_0}) it follows, that for the fixed
direction of the initial neutrino propagation, i.e. for $\cos
\theta =-1$, the cross section is
\begin{equation}\label{eq72}
  {\sigma_{{n{=}0}}}_{{\big |}_{ p'_{0}=m',\cos\theta=-1}}
  =\frac{eBG^2}{\pi}
  2\alpha^2(1-S)\frac{\Delta+\varkappa}{\sqrt{(\Delta+\varkappa)^2-m^2}}.
\end{equation}
For the neutron matter totally   polarized  parallel to the
magnetic field vector, $S=1$, the cross section is vanish. The
result of eq. (\ref{eq72}) coincides with one of
\cite{BhaPal02_1}.

The cross section in the strong magnetic field $B=B_{cr}$,
normalized to the cross section in the field-free case, calculated
with use of the exact eq. (\ref{cross_strong}) is shown in Fig.7.
Note also that, as it can be seen from the Fig.7, the cross
section for $\cos \theta =1$ and $S=-1$ is also rather small. This
is a consequence of smallness of the value $(\alpha-1)$ because
the cross section in this case is proportional to
$(1-\alpha)^{2}<0.1$. The neutrino energy is chose to be
$\varkappa = 10 \ MeV$, that is why the effects of the proton
recoil motion cannot be screened.

In Figs.8 (a,b,c) we plot the dependence of the cross section on
the strength of strong magnetic fields for the different intervals
within $B_{cr}\leq B <B'_{cr}$ in the case of unpolarized ($S=0$)
neutrons for the initial neutrino energy $\varkappa= 10 \ MeV$.
The super-strong magnetic field $B\geq B'_{cr} \ (B^{*}\sim 4.6)\
(B^{*}=\log {{B \over B_{0}}}$, where $B_{0}={m^2 \over e}= 4.41
\times 10^{13} \ G)$ is also included in the panel (c). The solid
curves correspond to $\cos \theta =1$, the dashed curve
corresponds to $\cos \theta = -1$. The is a fall in the cross
section for $\cos \theta =1$ in the magnetic field $B=B'_{cr}$
because in the strong magnetic field $B_{cr}\leq B <B'_{cr}$ the
cross section is zero only for one set of the neutron spin number
$s_n$ and $\cos \theta$, i.e. for the case when $(s_n=1,\ \cos
\theta =-1)$, whereas in the super-strong magnetic field $B\geq
B'_{cr}$ the cross section is zero for the two cases $(s_n=1,\
\cos \theta =-1)$ and $(s_{n}=-1,\ \cos \theta =1)$.

The analogous dependence of the cross sections on the strength of
the magnetic field for the totally polarized neutrons with $S=-1$
and $S=1$ are plotted in Figs.9 (a,b,c) and Figs.10 (a,b,c),
respectively. The super-strong magnetic field $B\geq B'_{cr} \
(B^{*}\sim 4.6)$ is also included in the panels (c) of Figs.9 and
10.

For $S=-1$ the cross section is small if the neutrino propagates
parallel to the magnetic field ($\cos \theta =1$) within the
interval $B_{cr}\leq B <B'_{cr}$ (see Figs.9 (a,b)) because the
cross section is proportional to $(1-\alpha)^{2} < 0.1$. In the
super-strong magnetic field $B\geq B'_{cr} \ (B^{*}\sim 4.6)$
(Fig.9 (c)) the cross section is exactly zero as we have already
discussed above. For $S=1$ in the case of $\cos \theta =-1$ for
the whole range $B \geq B_{cr}$ (Figs.10 (a,b,c)) the cross
section is also zero.

The shown in Figs.7-10 plots for the cross section in the magnetic
field disagree with the corresponding plots for the cross section
in the strong field range ($B\geq B_{cr}$) shown in Fig.1 of
\cite{BhaPal02_1}. The contradictions disappear if the solid and
dashed curves in Fig.1 of \cite{BhaPal02_1} are replaced. There is
also no rapid increase of the cross section in the field $B\geq
B_{cr}$ for the case of $\cos \theta =1$ and $S=-1$, contrary to
what is shown in the first panel of Fig.1 of \cite{BhaPal02_1}.

%\newpage

\subsection{Cross section in weak magnetic field}

In the case of weak magnetic fields $B< B_{cr}$ many Landau levels
become available for the electron so that the electron can have
non-zero momentum $p_{\perp}=\sqrt{2\gamma n}$ in the transverse
plane. The maximum allowed value for $n$ is given by
(\ref{n_max}). In the calculations of the cross section in the
presence of a weak magnetic field we also expand the
$\delta$-function (see eq. (\ref{delta_exp})) and perform the
summation over the proton Landau number $n'$ up to infinity. The
particular contribution to the cross section from the partial
process with the electron at the lowest Landau level ($n=0$) has
been already discussed in the previous Section 2.3. Therefore, we
derive now  the fraction $\sigma _{n\geq 1}$ of the total cross
section that is the sum of the corresponding contributions from
the excited electron Landau levels with $n\geq 1$. The final
result for the cross section can be expressed as
\begin{equation}\label{cross_tot}
  \sigma_\text{tot}=\sigma_{n{=}0}+\sigma_{n{\geq}1}.
\end{equation}

Putting the general expression for the squared matrix element
(\ref{mm}) to (\ref{cross}), then expanding over $\frac{p_3'}{m'}$
and performing summation over $n'$ (see the previous subsection),
we obtain to the first order in the proton recoil motion
\begin{multline}\label{cross_w_n}
  \sigma_{n{\geq}1}=\frac{eB G^2}{2\pi}
  \sum_{n=1}^{n_{max}}\sum_{i=1,2}
{1 \over
  {\Big|{p_{3}^{(i)}
  \over p_{0}^{(i)}}-{p_{3}'^{(i)}
 \over {{p}}_{0}'^{(i)}}\Big|}}
  \Bigg[1+3\alpha^2+2\alpha (1-\alpha)\frac{p_3'^{(i)}}{m'}
  (1+\cos\theta) +
  2S\alpha(1+\alpha)\cos\theta \\
    +2(1+\alpha)^2\frac{\gamma n}{p_0^{(i)}
  m'}(1+S\cos\theta)\Bigg].
  \end{multline}
In the limit of non-moving proton $(p'_0=m')$ the contribution to
the cross section for $n\geq 1$ is
\begin{multline}\label{cross_w_n_1}
  {\sigma_{n{\geq}1}}_{{\big |}_{p_0'{=}m'}} = \frac{eB G^2}{\pi}
  \Big[1+3\alpha^2+2S \alpha (1+\alpha)\cos\theta\Big] \\
  \times\sum_{n=1}^{n_{max}}\frac{\Delta+\varkappa}{\sqrt{(\Delta+\varkappa)^2-m^2-
  2\gamma n}}.
\end{multline}
Summing this result with one of eq. (\ref{cross_n_0_n'_0_p'_0}),
we get the result of ref.\cite{BhaPal02_1} for the total cross
section in the case of weak magnetic field (the proton recoil
motion is neglected here)
\begin{multline}\label{cross_total}
  \sigma_{\text{tot}{\big |}_{p'_0{=}m'}} = \frac{eB G^2}{2\pi}
  \sum_{n=0}^{n_{max}}\bigg\{g_n\Big[1+3\alpha^2+2S\alpha
  (1+\alpha)\cos\theta\Big]\\
  +\delta_{n,0}\Big[(1-\alpha^2)\cos\theta+2S\alpha
  (1-\alpha)\Big]\bigg\}
  \frac{\Delta+\varkappa}{\sqrt{(\Delta+\varkappa)^2-m^2-
  2\gamma n}}.
\end{multline}

As it follows from (\ref{cross_w_n_1}) and (\ref{cross_total}),
the cross section has several resonances (see also
\cite{Rou97,BhaPal02_1}). Similar resonances in the probability of
the direct beta-decay of the neutron in the magnetic field was
first discovered in \cite{Kor64,TerLysKor65}. In our case the
resonances appears, for the given neutrino energy $\varkappa$ and
magnetic field strength $B$, each time when the final electron
energy $p_0$ is exactly equal to one of the allowed ($n\leq
n_{max}$) "Landau energies" $\tilde p_{\perp}=\sqrt{m^2 + 2\gamma
n}$,
\begin{equation}\label{res}
  p_0=\varkappa + \Delta=\sqrt{m^2 + 2\gamma n}.
\end{equation}

In Figs.11 (a,b,c) we plot the cross section as a function of $B$
(in the range of not very strong magnetic fields, $B\leq B_{cr}$)
for the three different neutrino energies $\varkappa=30 \ MeV, 10\
MeV$ and $\varkappa\ll m$. Obviously, the similar resonance
behavior appears in the cross section as a function of the
neutrino energy in a given fixed magnetic field. The number of
resonances, which is equal to the number of terms in the sum of
eq. (\ref{cross_w_n_1}), increases with the increase of the
neutrino energy for a given $B$. The cross section, calculated
without effects of the proton recoil motion, goes to infinity in
the resonance points. However, if we plot the cross section with
use of eqs. (\ref{cross_strong}) and (\ref{cross_w_n}), which
accounts for the proton motion, then the infinitely high spikes
smooth out.

\subsection{Cross section in the absence of magnetic field}

The inverse beta-decay in the absence of a magnetic field was
considered before by many authors (see, for instance,
\cite{TubSch75,Bru85}). To the best of our knowledge, the
correlation between the neutron polarization and the direction of
the neutrino propagation for the scattering (V-A)-interaction
process was derived for the first time in ref.\cite{Ker61}. The
result for the cross section $\nu_e+n\rightarrow e+p$ in the
absence of the magnetic field is
\begin{equation}\label{cross_vac}
  \sigma_0=\frac{G^2}{\pi}\left[1+3\alpha^2+2\alpha S_{n}(1+\alpha)
  \cos \theta \right](\Delta+\varkappa)
  \sqrt{(\Delta+\varkappa)^2-m^2}.
\end{equation}
We now demonstrate, following the similar procedure described in
\cite{Kor64,TerLysKor65}, how in the limit of vanishing magnetic
field $B\rightarrow 0$ the result in eq. (\ref{cross_total})
reduces to the one of (\ref{cross_vac}).

When the field is switching off the maximum number of the Landau
level $n_{max}$ is increasing to infinity, however the product
$eBn$ remains constant,
\begin{equation}\label{lim}
    \lim_{\gamma\rightarrow 0, n\rightarrow \infty}\gamma n=
    \frac{(\Delta+\varkappa)^2-m^2}{2}.
\end{equation}
In this limit the summation over $n$ we can replace by integration
using the relation (see, for example, \cite{Kor64})
\begin{equation}\label{sum}
  \sum_{n=0}^N f(n)=\int\limits_0^N f(x)dx+\frac{f(N)+f(0)}{2}+
  \int_0^N Q(x)f''(x)dx,
\end{equation}
where the value of the last term can be estimated as
\begin{equation}\label{int}
  \int_0^n Q(x)f''(x)dx \leqslant\frac{f'(n)-f'(0)}{8}.
\end{equation}
In the sum over $n$ in (\ref{cross_total}) the contribution of the
lowest Landau level is diminishing in comparison with the
contributions of the exited Landau levels $n>0$. For the
estimation of the former we use
\begin{multline} \label{sum_int}
  \lim_{\gamma\rightarrow 0}\gamma\sum_{n=0}^{n_{max}}
  \frac{1}{\sqrt{(\Delta+\varkappa)^2-m^2-2\gamma n}} \\
  =\lim_{\gamma\rightarrow 0}\gamma\Bigg[\int\limits_0^{{n_{max}}}
  \frac{dx}{\sqrt{(\Delta+\varkappa)^2-m^2-2\gamma x}}
  +C\Bigg] \\
  =\lim_{\gamma\rightarrow 0}\gamma\bigg[\frac{1}{\gamma}
  \sqrt{(\Delta+\varkappa)^2-m^2}+C\bigg]=\sqrt{(\Delta+\varkappa)^2-m^2},
\end{multline}
where $C$ is a function proportional to $\gamma^{-1/2}$. Thus, in
the limit $B\rightarrow 0$ from (\ref{cross_total}) we get the
cross section of the process in the absence of a magnetic field.

\subsection{Effects of Anomalous Magnetic Moments of Nucleons}
When considering the influence of very strong magnetic fields on
the inverse beta-decay of a neutron one should be careful about
the effect of magnetic field on anomalous magnetic moments of a
neutron and proton. In particular, it is known (see, for instance,
\cite{BanRub91,GraRub01,MamGoud04}) that the interplay between
anomalous magnetic moments of the neutron and proton shifts the
masses of these particles. These effects are important only for
the super-strong magnetic fields, when  the corresponding shift of
the electron energy due to the electron anomalous magnetic moment
is vanishing \cite{TerBagDorJETP69,Schw88} (see also in
\cite{GraRub01}). As a result, the neutron becomes stable in the
presence of magnetic fields with the strength $B\geq 1.5 \times
10^{18} \ \ G$. On the other hand, the proton becomes unstable in
respect to the the inverse beta-decay $ p\rightarrow n + e^{+}
+\nu_e$ if the magnetic field is increased past the strength
$B\geq 2.7 \times 10^{18} \ \ G$. Therefore, in this section in
order to complete the relativistic theory of the neutron inverse
beta-decay in the super strong magnetic field, we discuss in some
detail the possible effect of the nucleons anomalous magnetic
moments interactions with a magnetic field.

 The energy of the moving proton and the neutron at rest in a magnetic field, with the
contributions from the anomalous magnetic moments interaction
being accounted for, are given respectively by
\begin{gather}\label{energy_amm}
  p'_0=\sqrt{\left(\sqrt{{m'}^2+2eBn'}-s'k_{p}B\right)^2+{p'_{3}}^2},
\\
p^{n}_0=m_n - s_{n}k_{n}B,
\end{gather}
where the values of the proton and neutron anomalous magnetic
moments
\begin{gather}\label{p_n_amm}
  k_p=\frac{e}{2m'}(\frac{g_p}{2}-1)\\
  k_n=\frac{e}{2m_n}\frac{g_n}{2},
\end{gather}
are determined by the Lande's $g-$factors: $g_p=5.58$,
$g_n=-3.82$.

Taking into account these modified expressions for the proton and
neutron energies, we can repeat all the described above
calculations of Section 2.1 applying the substitutions

\begin{gather}\label{Mass_AMM}
  m'\rightarrow{m'}^*=m'-k_pB,\\
 m_{n} \rightarrow{m_n}^*=m_n-s_{n}k_{n}B.
\end{gather}

Note that in the super-strong magnetic filed $B\geq B'_{cr}$ there
is the only one spin state for the proton with $s'=+1$.

The law of energy conservation (\ref{energy}) shows that in the
super-strong magnetic field there is a range of the neutron matter
polarization $S$ for which the matter becomes transparent for
neutrinos. From  (\ref{energy}) we get:
\begin{equation}\label{eq_B_proh}
    m_n-s_{n}k_{n}B+\varkappa\geq m+m'-k_{p}B.
\end{equation}
Therefore, the process $\nu_e + n\rightarrow e+p$ is forbidden if
$(Sk_{n}-k_{p})>0$ and the magnetic field exceeds the value of
$B_{forb}$:

\begin{equation}\label{B_forb}
  B_{forb}=\frac{\Delta+\varkappa-m}{Sk_{n}-k_{p}}.
\end{equation}
Note that this forbidding effect appears for nearly maximum
neutron matter polarizations against the magnetic field, $-1\leq S
< k_{p}/k_{n}\approx -0.94 $. The values of $B_{forb}$ for
different neutrino energies in case of maximum neutron spin
polarization $S=-1$  are
\begin{equation}\label{}
B_{forb}\approx 8.5 \times 10^{19} G, \ \ \varkappa_{max}=30 \
MeV,
\end{equation}
\begin{equation}\label{}
B_{forb}\approx 3.0 \times 10^{19} G, \ \ \varkappa_{max}=10 \
MeV,
\end{equation}
\begin{equation}\label{}
B_{forb}\approx 2.2 \times 10^{18} G, \ \ \varkappa_{max} \ll \ m.
\end{equation}

\section{Conclusions}
We have developed the relativistic theory of the inverse
beta-decay of the polarized neutron in a magnetic field. Effects
of the proton momentum quantization in the magnetic field have
been included. The obtained closed expression for the cross
section in the magnetic field exactly accounts for the proton
longitudinal and transversal motion. For the three ranges of the
magnetic field (which we call the super-strong magnetic field
$B\geq B'_{cr}$, the strong field $B_{cr}\leq B<B'_{cr}$, and the
weak field $B<B_{cr}$) we have calculated the cross section and
discussed its dependence on the neutrino energy and angle
$\theta$, as well as on the neutron polarization $S$.

In description of the proton we have used the the exact solution
of the Dirac equation in a magnetic filed. This enables us to get
the exact cross section in the case of the super-strong magnetic
field $B\geq B'_{cr}$ when the proton can occupy only the lowest
Landau level $n'=0$. We have shown that it is not correct to use
the cross section, derived under the assumption that the proton
wave function is not modified by the magnetic field, in the case
when only one or not  many Landau levels are opened  for the
proton even if the proton motion is neglected. From the obtained
expressions for the cross section in the strong and super-strong
magnetic filed it is clearly seen that
\begin{equation}\label{concl}
  \sum_{n'=0}^{\infty}\sigma(n,n')_
{{\big |}_{ p'_{0}=m'}} \neq \sigma(n,n')_{{\big |}_{ n'=0}},
\end{equation}
and even
\begin{equation}\label{concl}
  \sum_{n'=0}^{\infty}\sigma(n,n')_
{{\big |}_{ p'_{0}=m'}} \neq
\sum_{n'=0}^{n'_{max}}\sigma(n,n')_{{\big |}_{ n'=0}},
\end{equation}
if not  many Landau levels $n'$ are available. Thus we conclude
that the Landau quantization of the proton momentum have to be
accounted for not only the super-strong magnetic field, but even
for lower magnetic fields when not too many Landau levels are
opened for the proton.

We should also like to point out here that it is not possible to
use the expression of the cross section, derived for the strong
magnetic field ($B_{cr}<B\leq B'_{cr}$), in the case of the
super-strong magnetic field ($B\geq B'_{cr}$) and also for lower
magnetic fields ($B\leq B'_{cr}$) when only a few Landau levels
for the proton are available.

We have shown that in the case of the total neutrons polarization
($S=\pm1$) the cross section is exactly zero in the super-strong
magnetic filed if $S\cos\theta=-1$, i.e. in the two cases: $1) \
S=1,\ \cos \theta =-1$, and $2) \ S=-1,\ \cos \theta =1$. Thus, in
the super-strong magnetic field the totally polarized neutron
matter is transparent for the neutrino propagating in the
direction opposite to the direction of the neutrons polarization.
In the case of the strong magnetic filed the cross section is
exactly zero if $S=1$ and $\cos \theta =-1$, that confirms the
result of ref.\cite{BhaPal02_1}. These asymmetries in the cross
section appear as a consequence  of the angular momentum
conservation and of the spin polarization properties of the
electron and proton being at the lowest Landau levels in the
magnetic field.

It should be noted that the developed relativistic treatment of
the cross section  can be applied to the other URCA processes with
two particles in the initial and final states. For instance,
similar calculations can be performed for the anti-neutrino
absorption process on the proton in the presence of a magnetic
field,
\begin{equation}\label{anti_nu}
  \bar\nu_e + p \rightarrow n  + e^{+}.
\end{equation}
The recent study of this process in strong-magnetic field without
account for the neutron recoil can be found in \cite{HuaQian04_2}.
The crossing symmetry makes it possible, with use of the matrix
element of the neutron inverse beta-decay, to write the matrix
element of the former process immediately. What remains to be done
is to change appropriately the phase volume of the process. With a
minor modification, the obtained above expressions for the cross
section of the neutron inverse beta-decay can be transformed for
the process $\bar\nu_e p \rightarrow n  e^{+}$ in a magnetic
field. For example, the obtained above expressions for the cross
section give the cross section of the process $\bar\nu_e p
\rightarrow n e^{+}$ if the signs of the values $\Delta$ and
$\alpha$ are changed to the opposite and also the substitution
$s_n \rightarrow s'$ is made.

%%%% Picture  %%%%%%%%%%%%%
\newpage

%\include{picture}
%%%%%%%%%%%%%%%%%%%%%%%%%%%%%%%%%%%%%%%%%%%%%%%%%%%%%%%%
\begin{figure}[]
  \centering
  \includegraphics[scale=0.6]{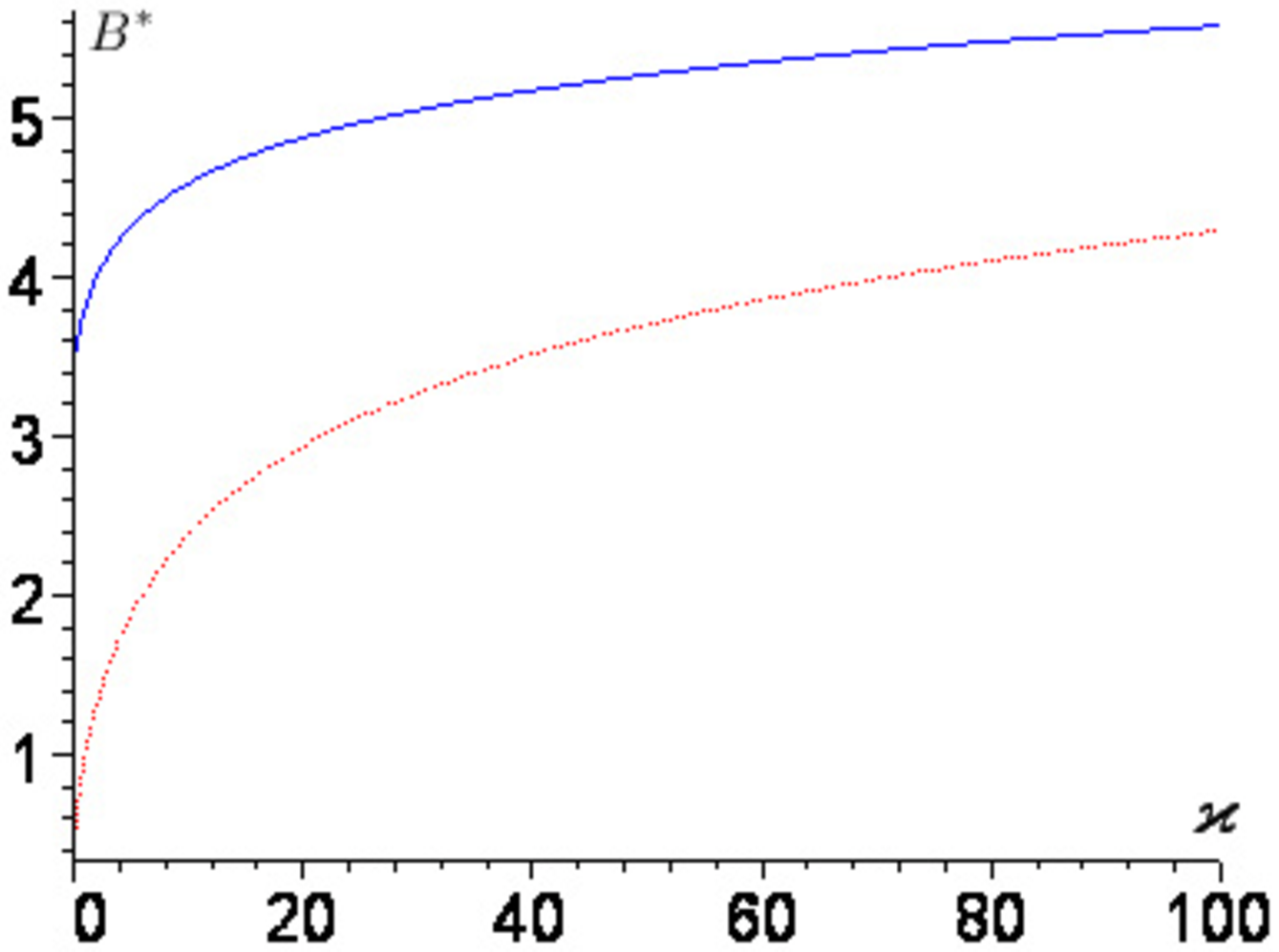}
   \caption{\small{Dependence of the electron critical magnetic field
$B_{cr}$ (dashed line) and the proton critical magnetic field
$B'_{cr}$ (solid line) on the initial neutrino energy $\kap \
(MeV$). The logarithmic scale is used: $B^{*}=\log {{B \over
B_{0}}}$, where $B_{0}={m^2 \over e}$. }}\label{fig1}
\end{figure}
%%%%%%%%%%%%%%%%%%%%%%%%%%%%%%%%%%%%%%%%%%%%%%%%%%%%%%%%%%%%%

%%%%%%%%%%%%%%%%%%%%%%%%%%%%%%%%%%%%%%%%%%%%%%%%%%%%%%%%%%%%%
\begin{figure}[]
\begin{center}
  \subfigure[]
  {\label{fig2a}
    \includegraphics[scale=0.55]{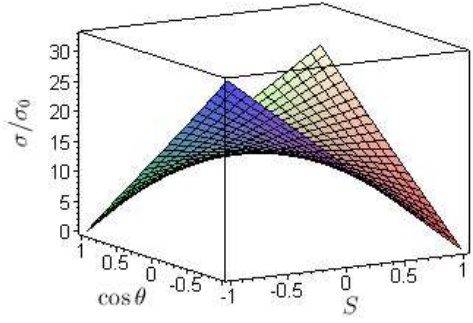}}
  %  \hspace{2cm}
  \subfigure[]
  {\label{fig2b}
  \includegraphics[scale=0.55]{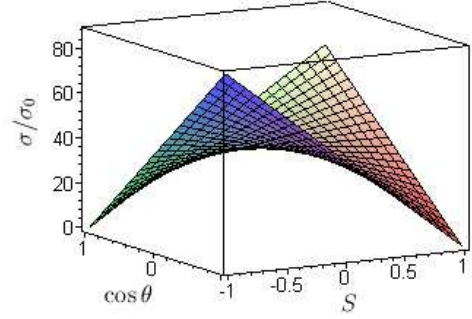}}
      \subfigure[]
  {\label{fig2c}
  \includegraphics[scale=0.55]{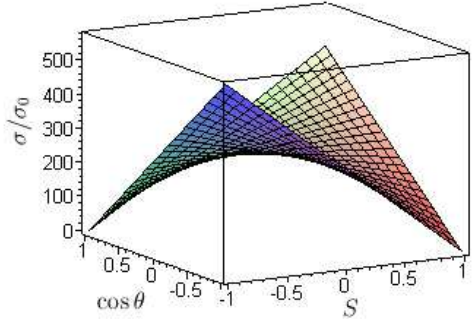}}
   % \hspace{2cm}
    \caption{\subref{fig2a}-\subref{fig2c}
    \small{The cross section $\sigma$ in super-strong
magnetic field $B=B'_{cr}$, normalized to the cross section
$\sigma _{0}$ in the field-free case, for neutrinos with energy of
$\kap =30 \ MeV$(a), $10\ MeV$ (b) and $\kap\ll m$ (c) as
functions of the direction of the neutrino momentum $\cos \theta$
and polarization
 of neutrons $S$.}}
\end{center}
\end{figure}

%%%%%%%%%%%%%%%%%%%%%%%%%%%%%%%%%%%%%%%%%%%%%%%%%%%%%%%%%%%%%%%%

%%%%%%%%%%%%%%%%%%%%%%%%%%%%%%%%%%%%%%%%%%%%%%%%%%%%%%%%%%%%%
\begin{figure}[]
  \centering
  \includegraphics[scale=0.9]{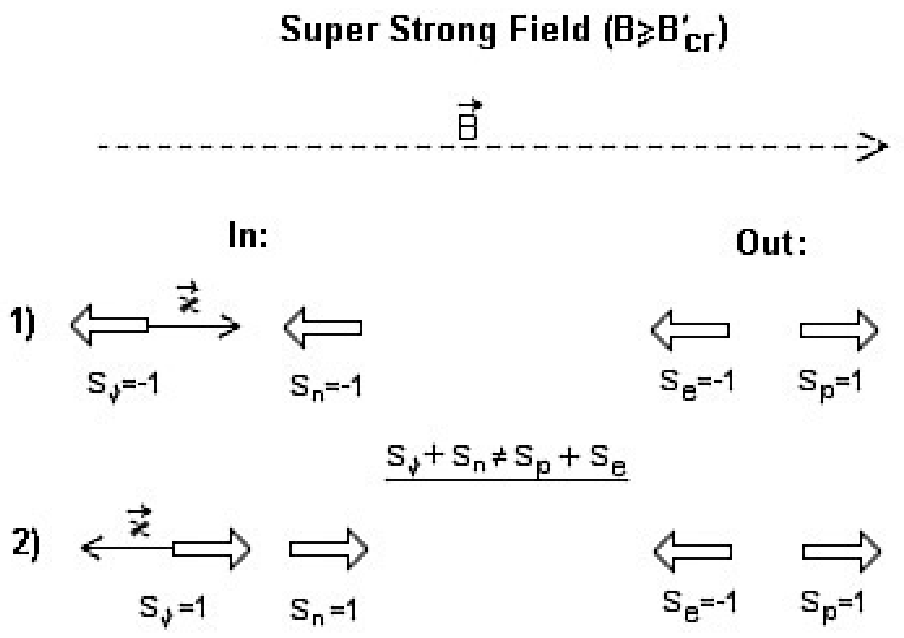}
   \caption{\small{Initial and final particles spin orientations for the two
directions of the neutrino propagation ($\cos \theta =\pm 1$) in
the super-strong magnetic field $B\geq B'_{cr}$. The broad arrows
represent the particles spin orientations, the solid arrows show
directions of the neutrino propagation, and the dashed arrow shows
the direction of the magnetic field vector. The cross section is
zero when the sum of the spin numbers of the initial particles
($s_{\nu}+s_n =\pm 2$) is not equal to the sum of the spin numbers
of the final particles ($s+s' =0$).}}\label{fig3}
\end{figure}

%%%%%%%%%%%%%%%%%%%%%%%%%%%%%%%%%%%%%%%%%%%%%%%%%%%%%%%%%%%%
\begin{figure}[]
\begin{center}
  \subfigure[]
  {\label{fig4a}
    \includegraphics[scale=0.55]{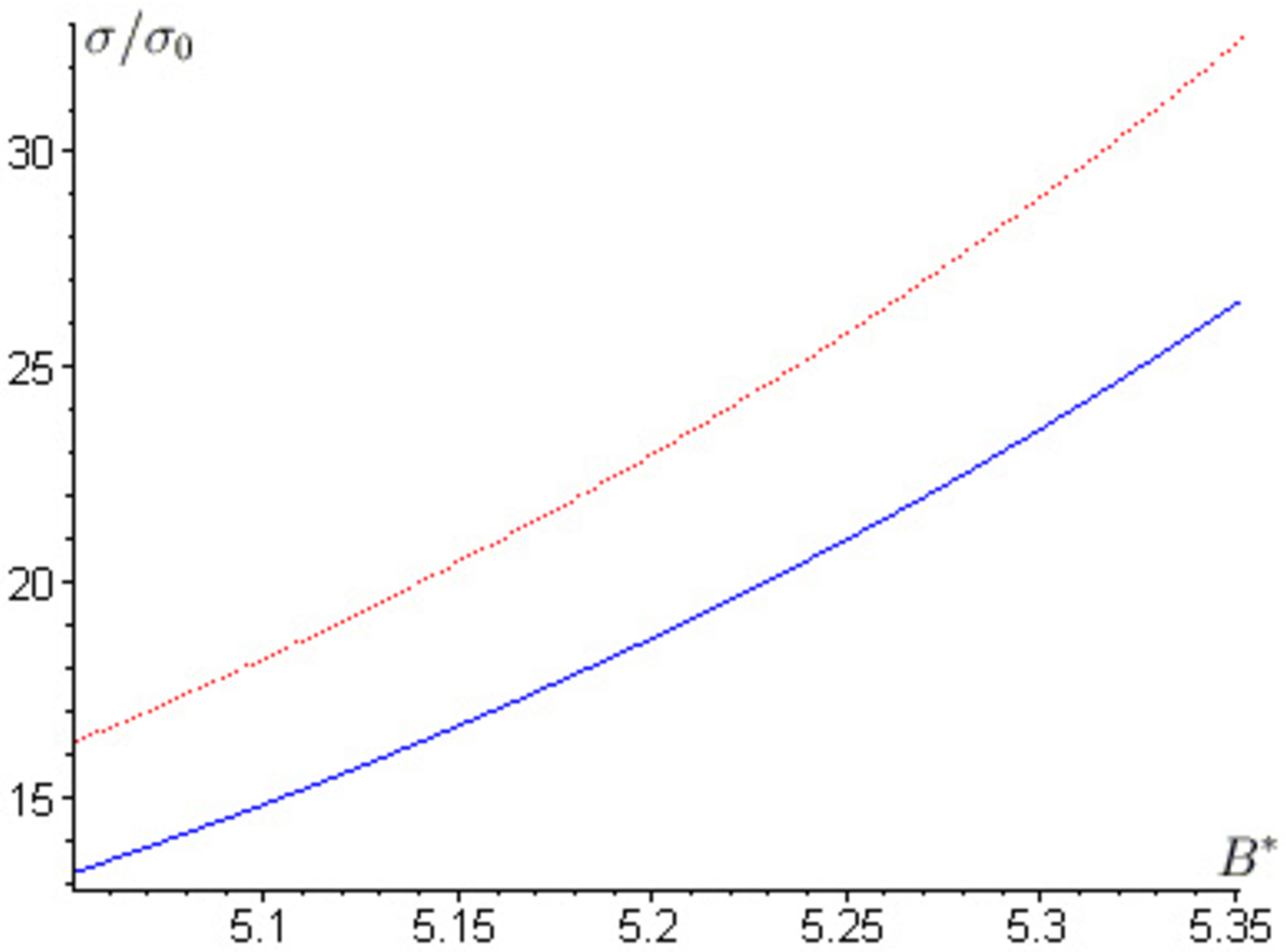}}
  %  \hspace{2cm}
  \subfigure[]
  {\label{fig4b}
  \includegraphics[scale=0.55]{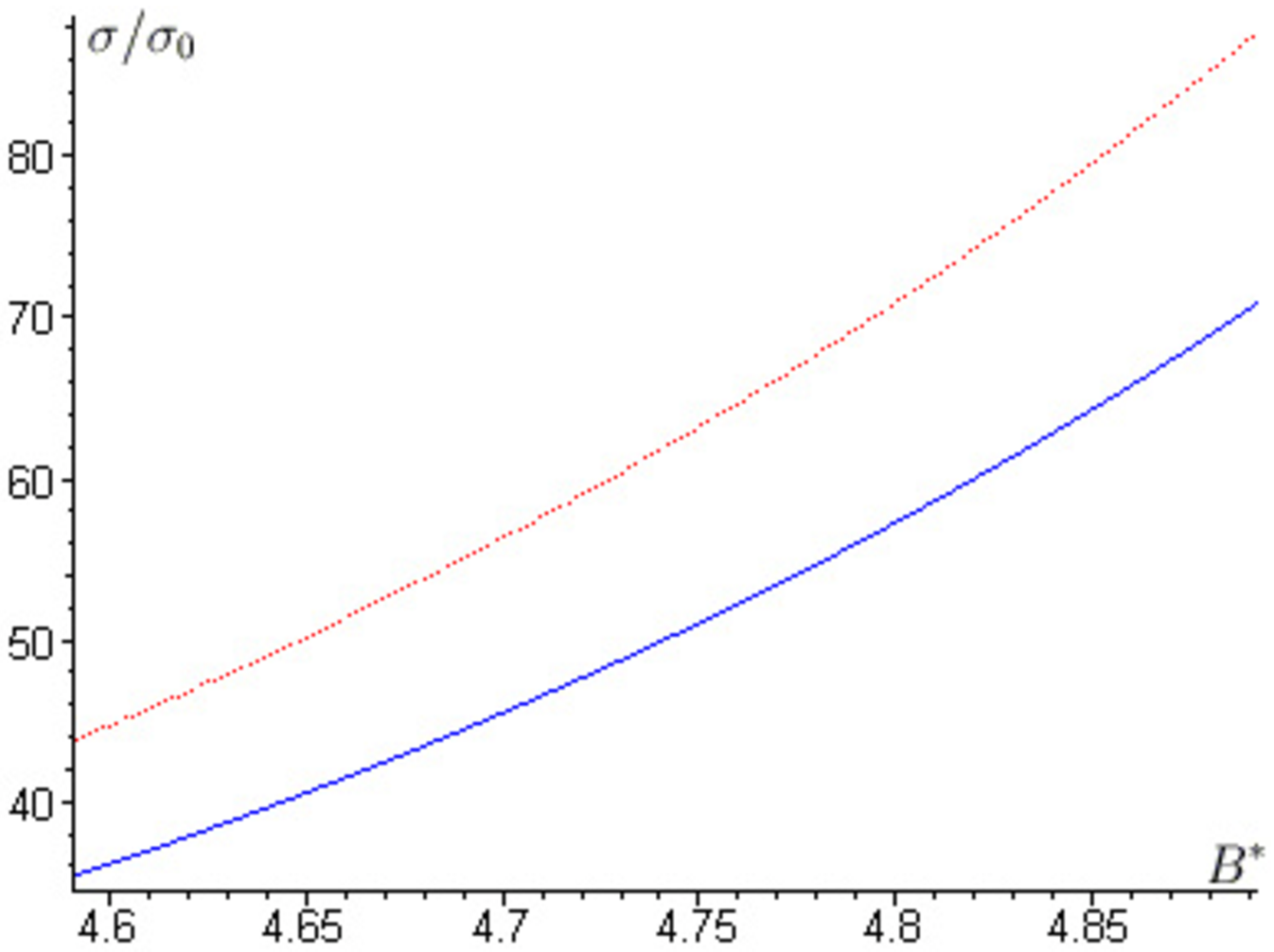}}
      \subfigure[]
  {\label{fig4c}
  \includegraphics[scale=0.55]{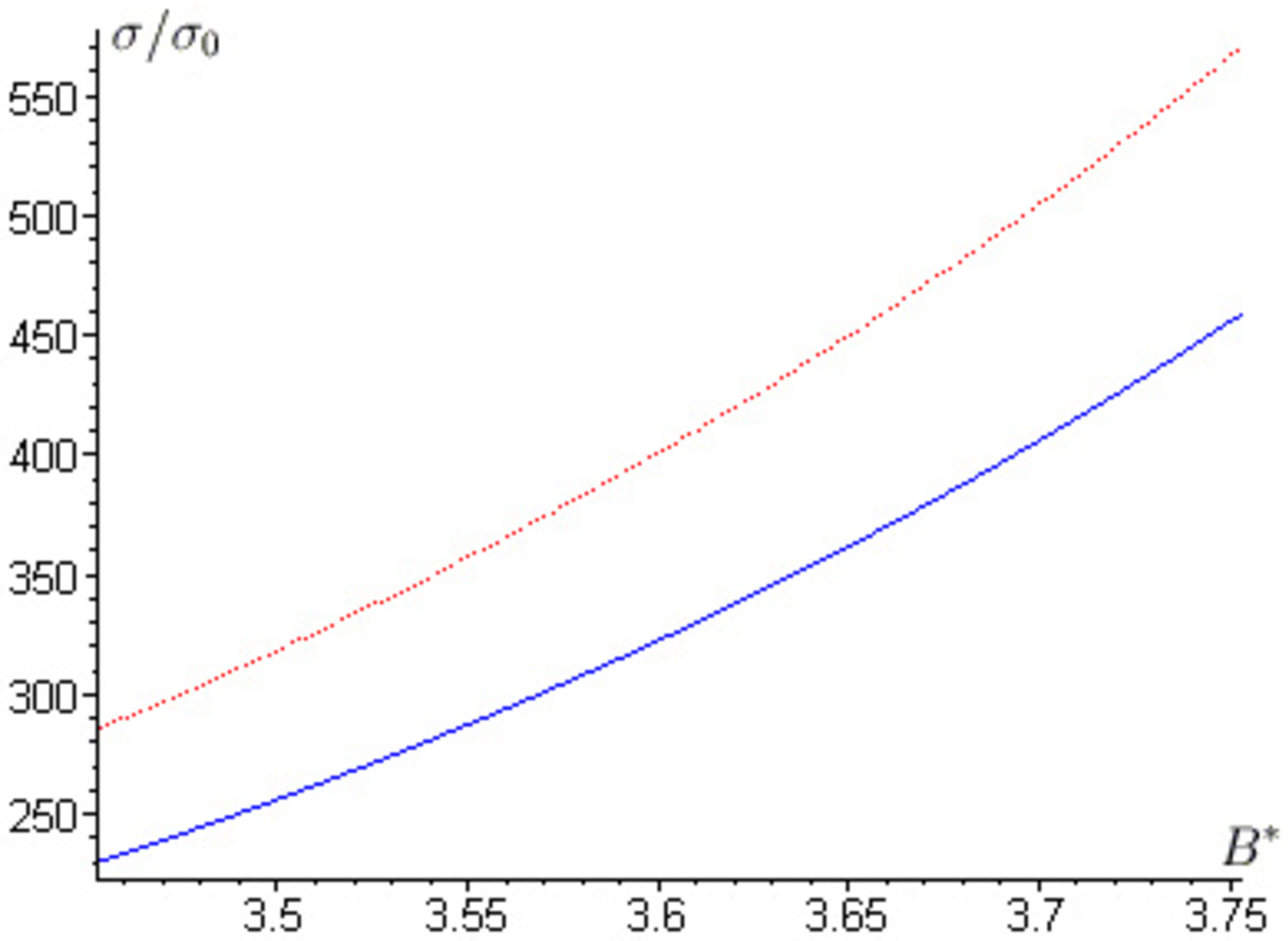}}
   % \hspace{2cm}
    \caption{\subref{fig4a}-\subref{fig4c}
    \small{
    The cross section in the super-strong magnetic field
$B\geq B'_{cr}$, normalized to the field-free case, for different
neutrino energies ($\kap=30 \ MeV$(a), $10\ MeV$ (b) and $\kap\ll
m$ (c)) in the case of unpolarized neutrons, $S=0$. The solid and
dashed lines correspond to the initial neutrino propagation
parallel ($\cos \theta=1$) and antiparallel ($\cos \theta =-1$) to
the magnetic field vector. The logarithmic scale is used:
$B^{*}=\log {{B \over B_{0}}}$, ($B_{0}={m^2 \over e})$.}}
\end{center}
\end{figure}
%%%%%%%%%%%%%%%%%%%%%%%%%%%%%%%%%%%%%%%%%%%%%%%%%%%%%%%%%%%%

%%%%%%%%%%%%%%%%%%%%%%%%%%%%%%%%%%%%%%%%%%%%%%%%%%%%%%%%%%%%
\begin{figure}[]
\begin{center}
  \subfigure[]
  {\label{fig5a}
    \includegraphics[scale=0.55]{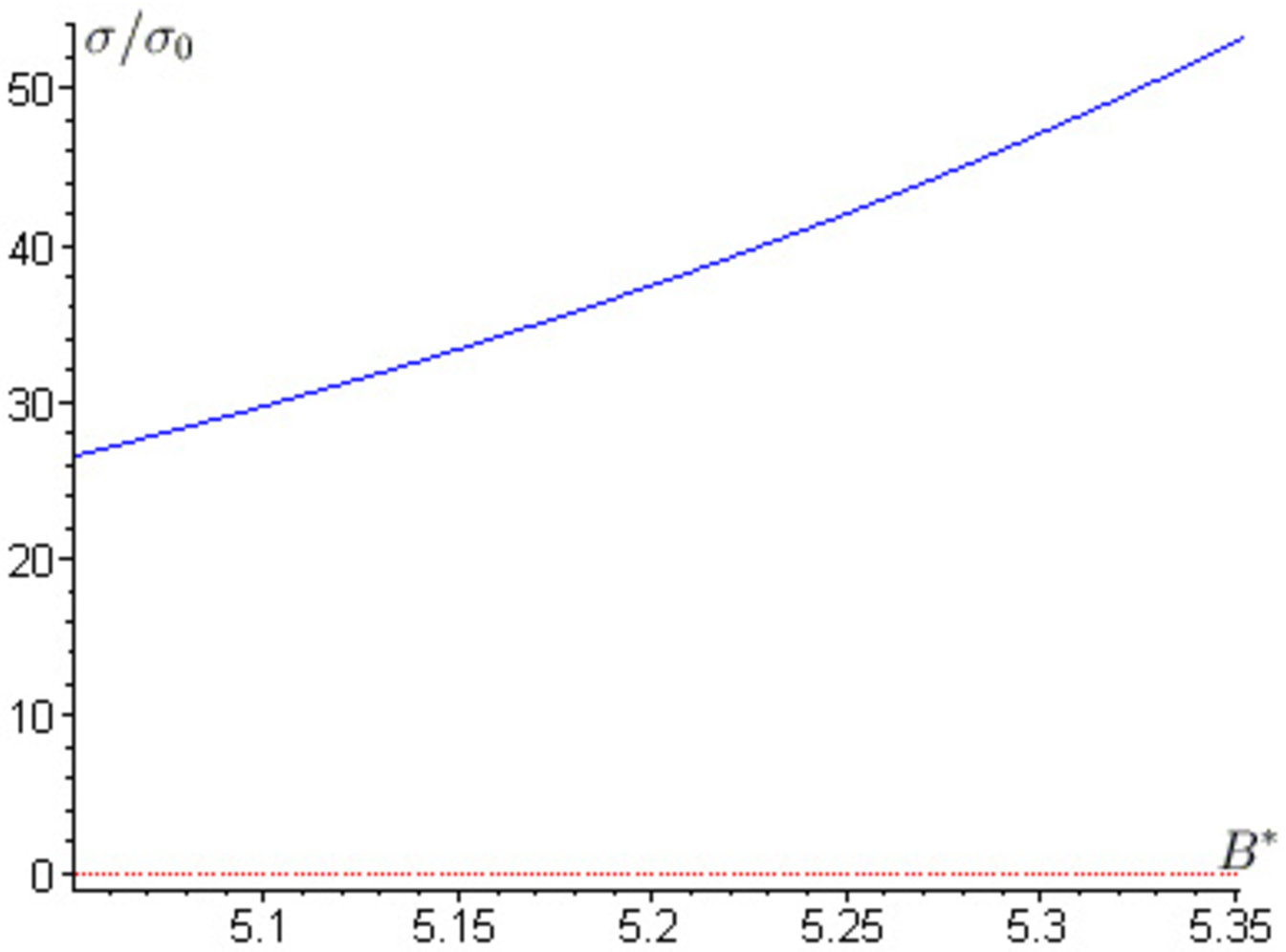}}
  %  \hspace{2cm}
  \subfigure[]
  {\label{fig5b}
  \includegraphics[scale=0.55]{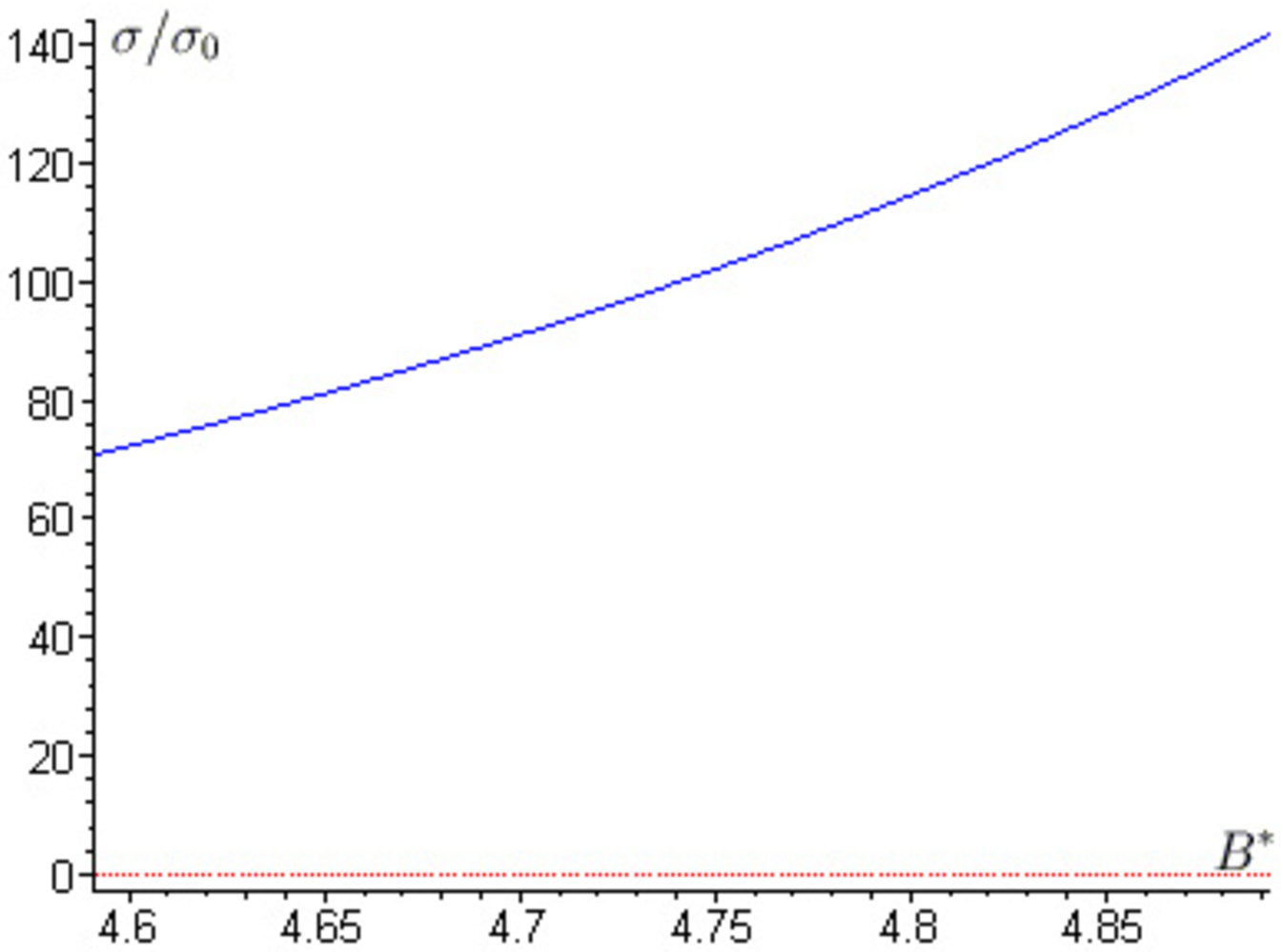}}
      \subfigure[]
  {\label{fig5c}
  \includegraphics[scale=0.55]{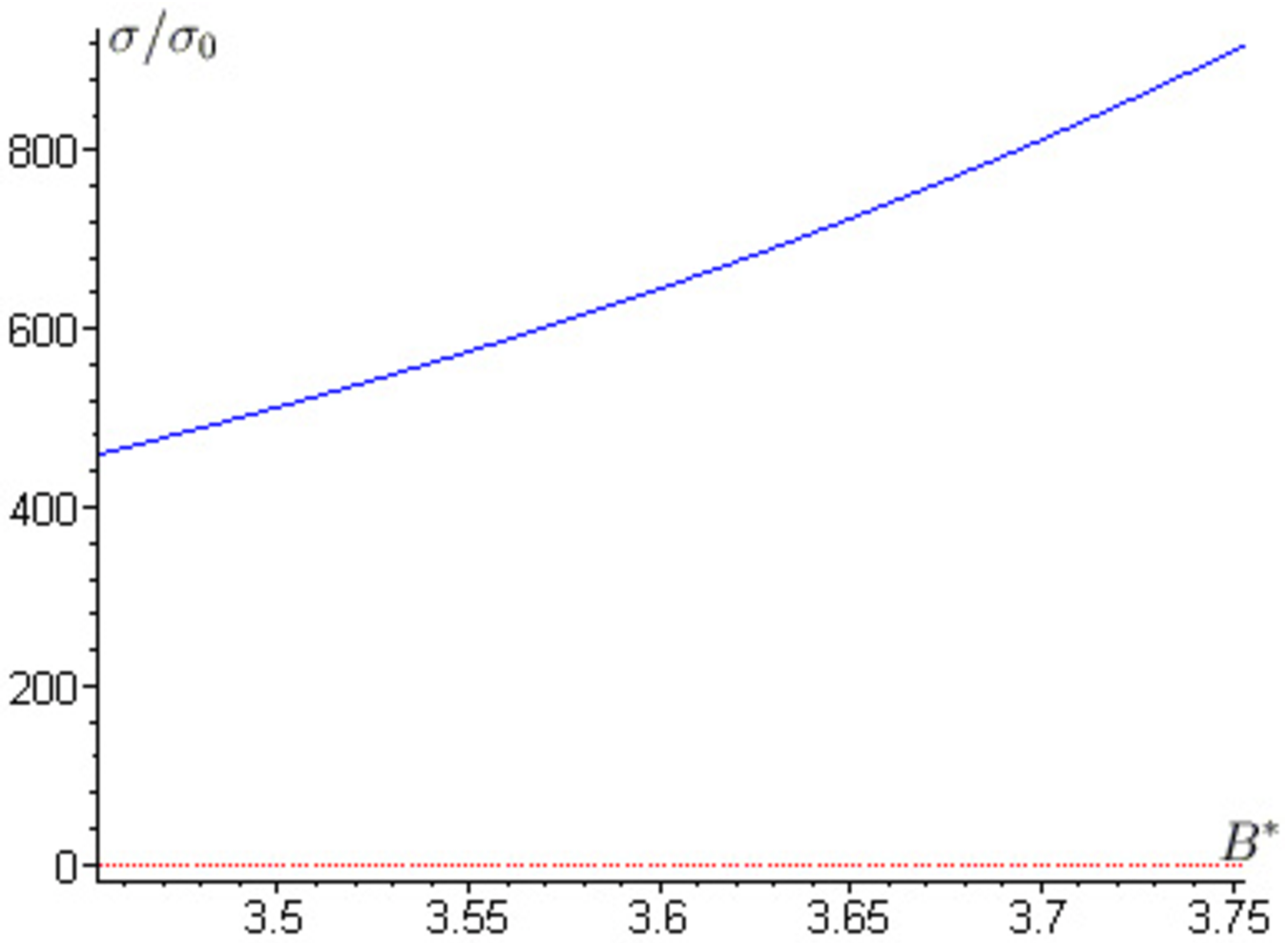}}
   % \hspace{2cm}
    \caption{\subref{fig5a}-\subref{fig5c} \small{
   The cross section in the super-strong magnetic field
    $B\geq B'_{cr}$, normalized
    to the cross section in the field-free case,
for different neutrino energies ($\kap =30 \ MeV$(a), $10\ MeV$
(b) and $\kap\ll m$ (c)) for neutrons totally polarized parallel
to the magnetic field vector ($S=1$). The solid and dashed lines
correspond to the initial neutrino propagation along ($\cos
\theta=1$) and against ($\cos \theta =-1$) the magnetic field
vector. The cross section for $\cos \theta =-1$ is exactly zero.
The logarithmic scale is used: $B^{*}=\log {{B \over B_{0}}}$,
where $B_{0}={m^2 \over e}$.}}
\end{center}
\end{figure}

%%%%%%%%%%%%%%%%%%%%%%%%%%%%%%%%%%%%%%%%%%%%%%%%%%%%%%%%%%%%
\begin{figure}[]
\begin{center}
  \subfigure[]
  {\label{fig6a}
    \includegraphics[scale=0.55]{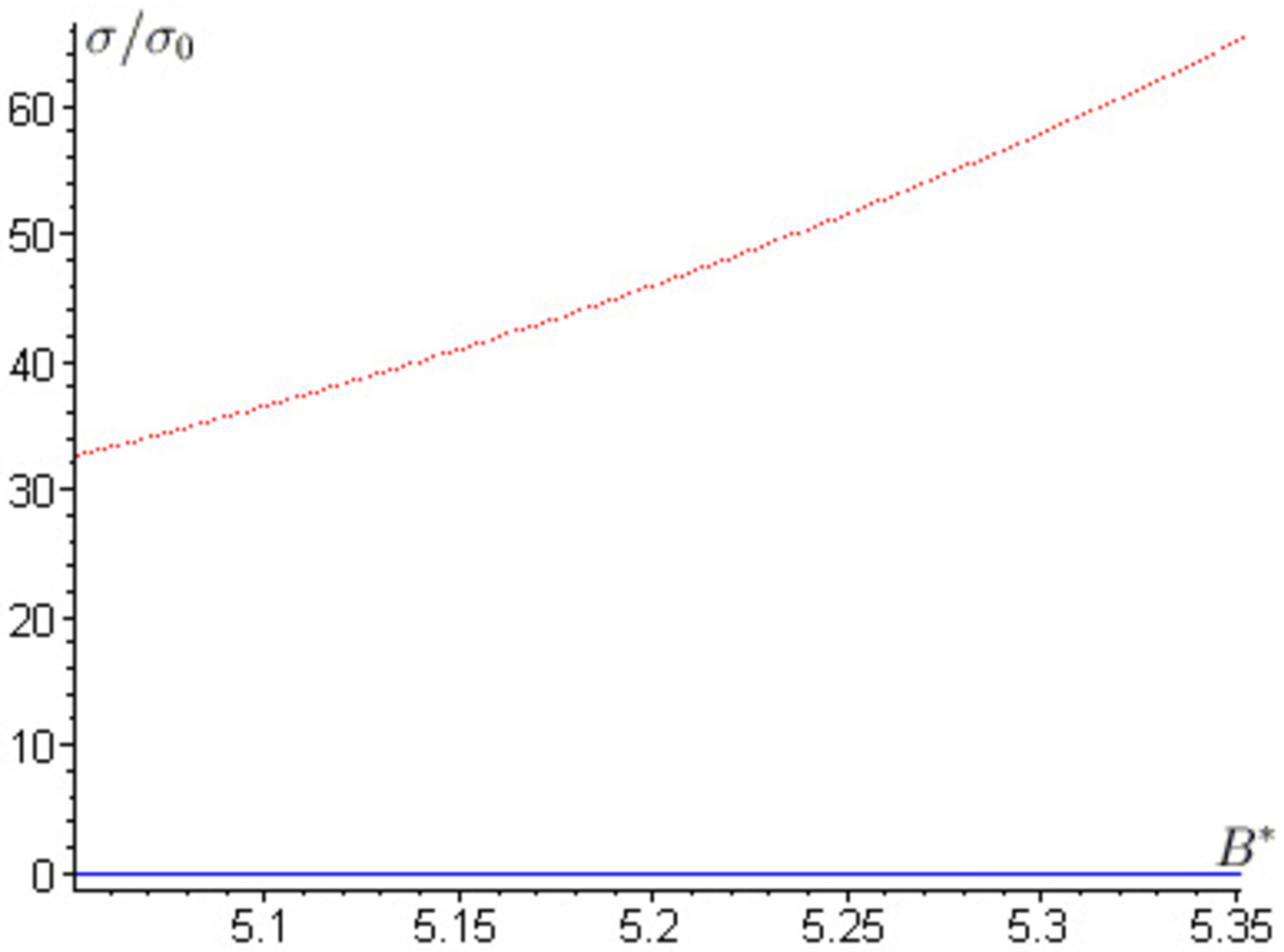}}
  %  \hspace{2cm}
  \subfigure[]
  {\label{fig6b}
  \includegraphics[scale=0.55]{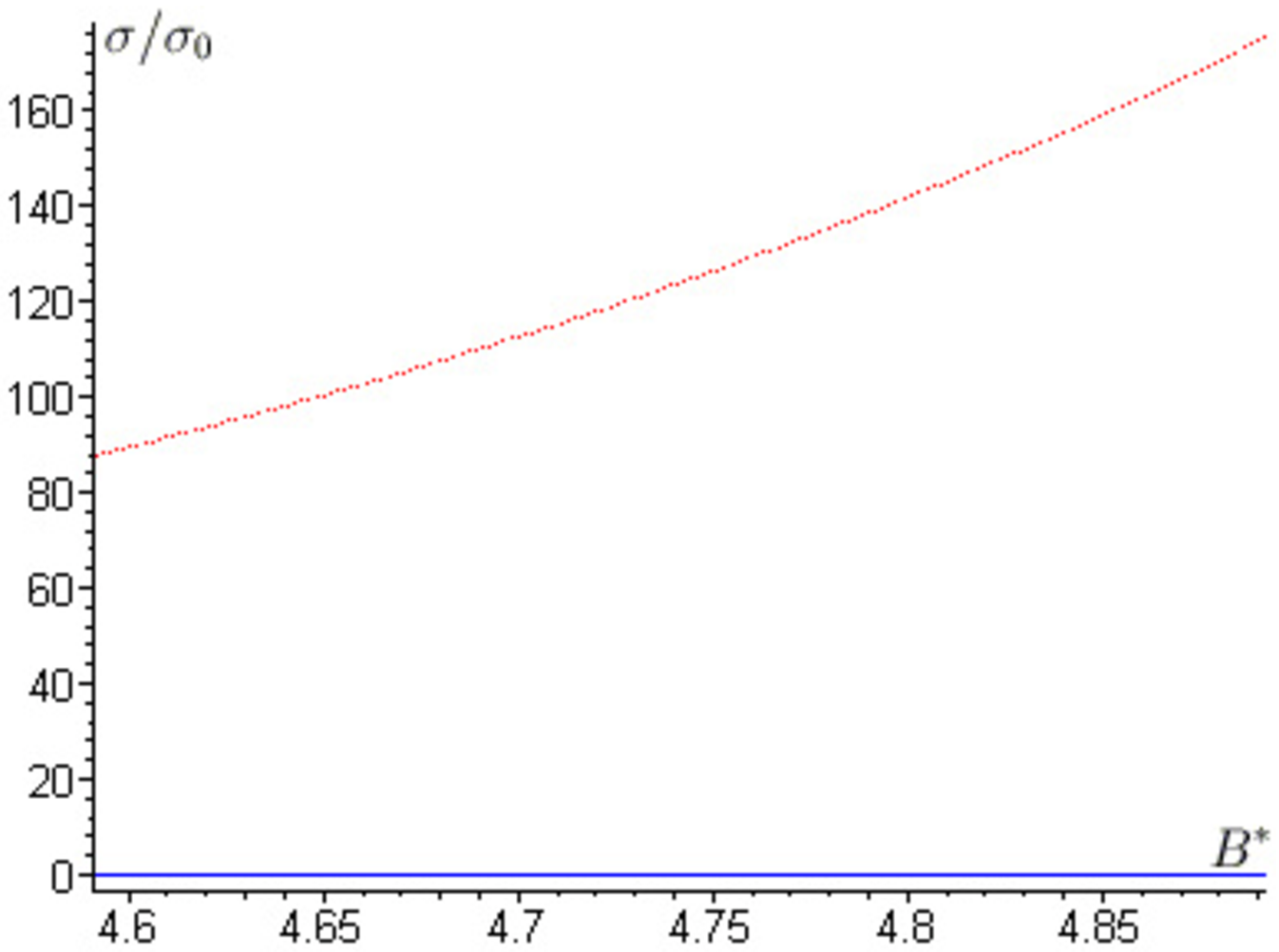}}
      \subfigure[]
  {\label{fig6c}
  \includegraphics[scale=0.55]{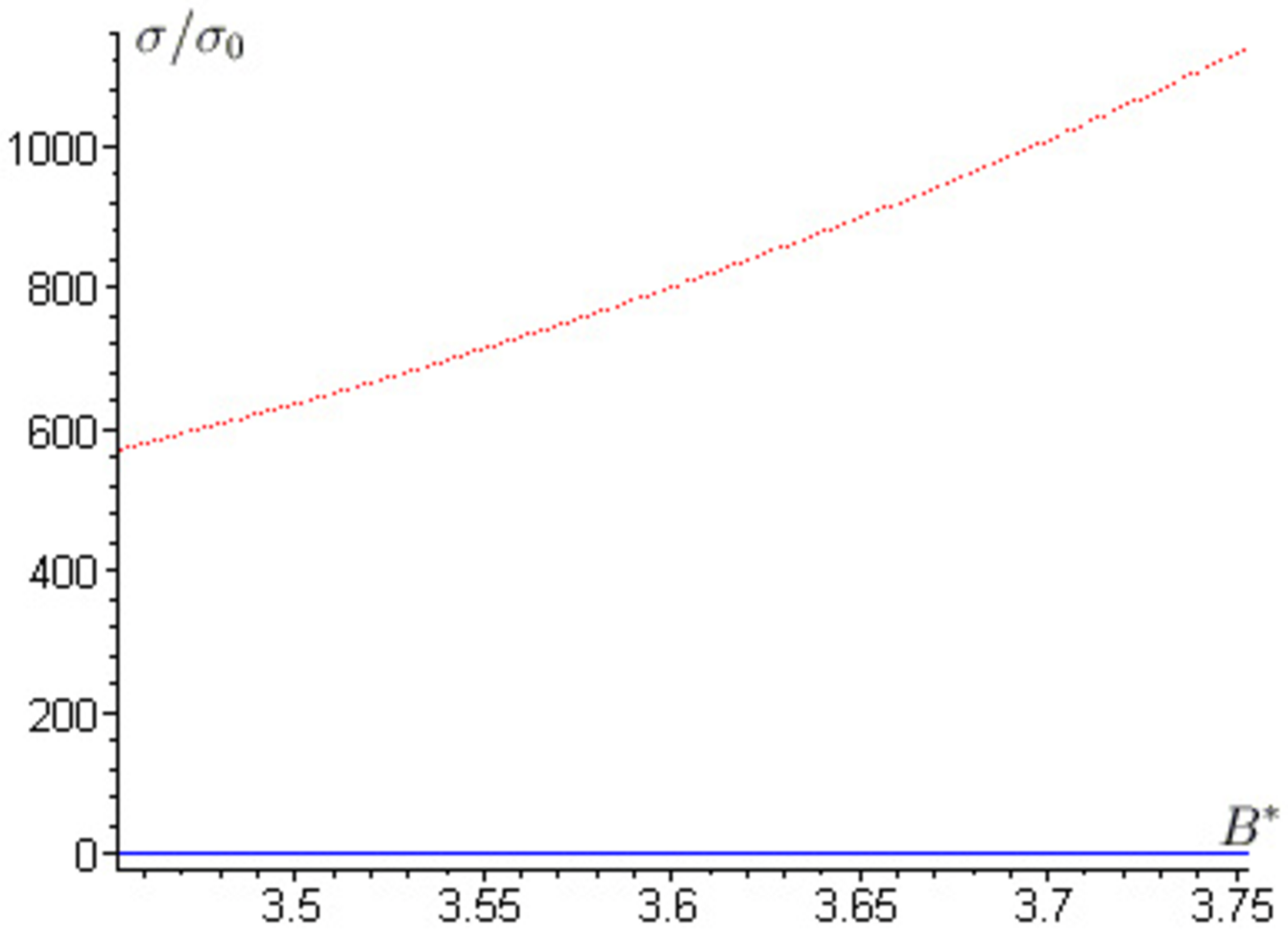}}
   % \hspace{2cm}
    \caption{\subref{fig6a}-\subref{fig6c}
    \small{The cross section in the super-strong magnetic field
    $B\geq B'_{cr}$, normalized
    to the cross section in the field-free case,
for different neutrino energies ($\kap =30 \ MeV$(a), $10\ MeV$
(b) and $\kap\ll m$ (c)) for neutrons totally polarized
antiparallel to the magnetic field vector ($S=-1$). The solid and
dashed lines correspond to the initial neutrino propagation along
($\cos \theta=1$) and against ($\cos \theta =-1$) the magnetic
field vector. The cross section for $\cos \theta =1$ is exactly
zero. The logarithmic scale is used: $B^{*}=\log {{B \over
B_{0}}}$, where $B_{0}={m^2 \over e}$.}}
\end{center}
\end{figure}
%%%%%%%%%%%%%%%%%%%%%%%%%%%%%%%%%%%%%%%%%%%%%%%%%%%%%%%%%%%%

%%%%%%%%%%%%%%%%%%%%%%%%%%%%%%%%%%%%%%%%%%%%%%%%%%%%%%%%%%%%%
\begin{figure}[]
  \centering
  \includegraphics[scale=0.6]{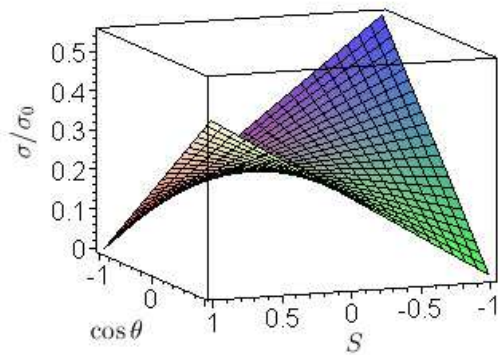}
   \caption{\small{The cross section in the strong magnetic field $B=B_{cr}$,
normalized to the cross section in the field-free case, for
neutrinos with energy of $10\ MeV$  as functions of the direction
of the neutrino momentum $\cos \theta$ and polarization $S$ of
neutrons. The cross section in the magnetic field in the case
$\cos \theta=-1, \ S=1$ is exactly zero, whereas the cross section
in the case $\cos \theta=1, \ S=-1$ is not zero, however it is
rather small because it is proportional to $(1-\alpha)^{2} <
0.1$.}}\label{fig7}
\end{figure}
%%%%%%%%%%%%%%%%%%%%%%%%%%%%%%%%%%%%%%%%%%%%%%%%%%%%%%%%%%%%

%%%%%%%%%%%%%%%%%%%%%%%%%%%%%%%%%%%%%%%%%%%%%%%%%%%%%%%%%%%%
\begin{figure}[]
\begin{center}
  \subfigure[]
  {\label{fig8a}
    \includegraphics[scale=0.55]{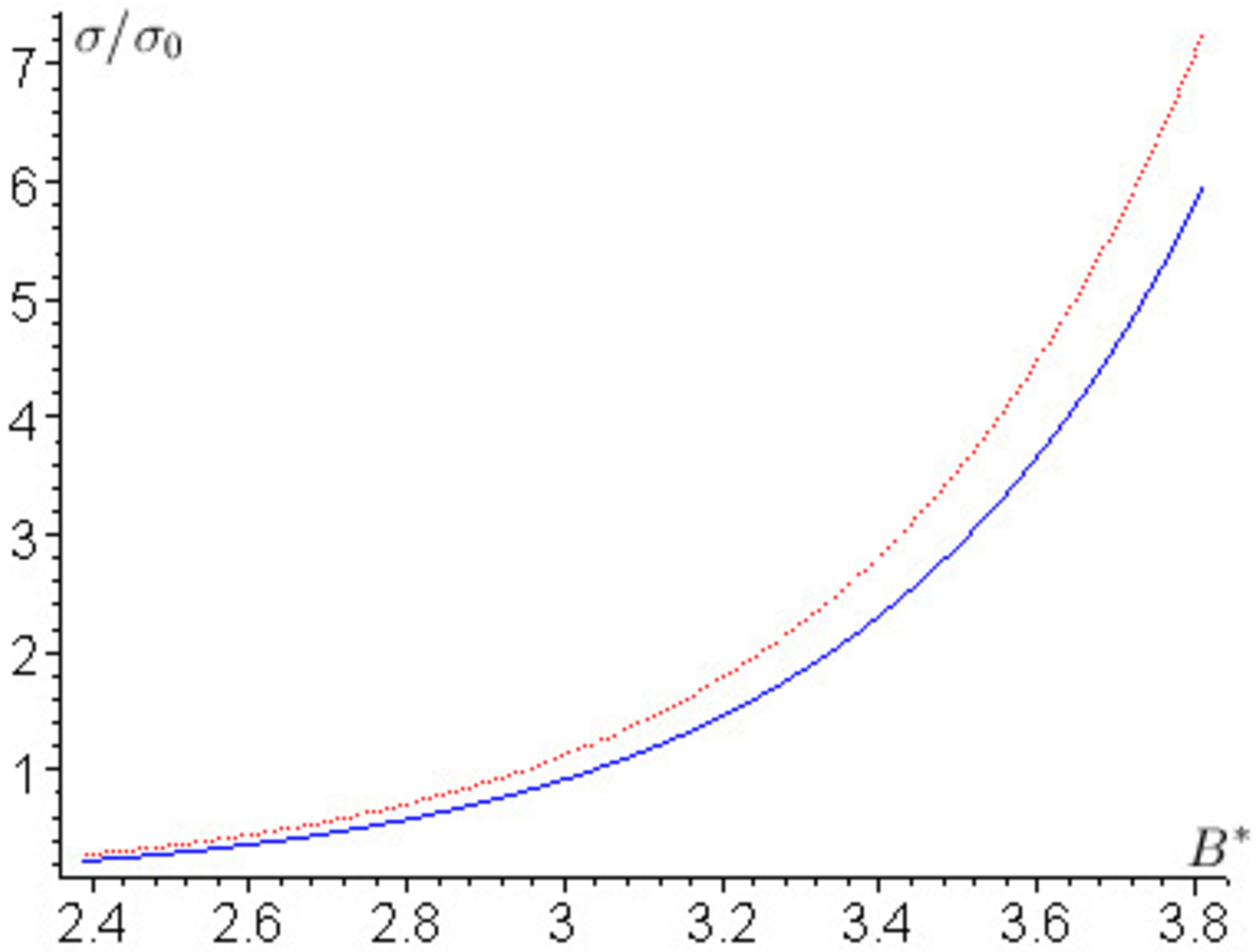}}
    \subfigure[]
  {\label{fig8b}
  \includegraphics[scale=0.55]{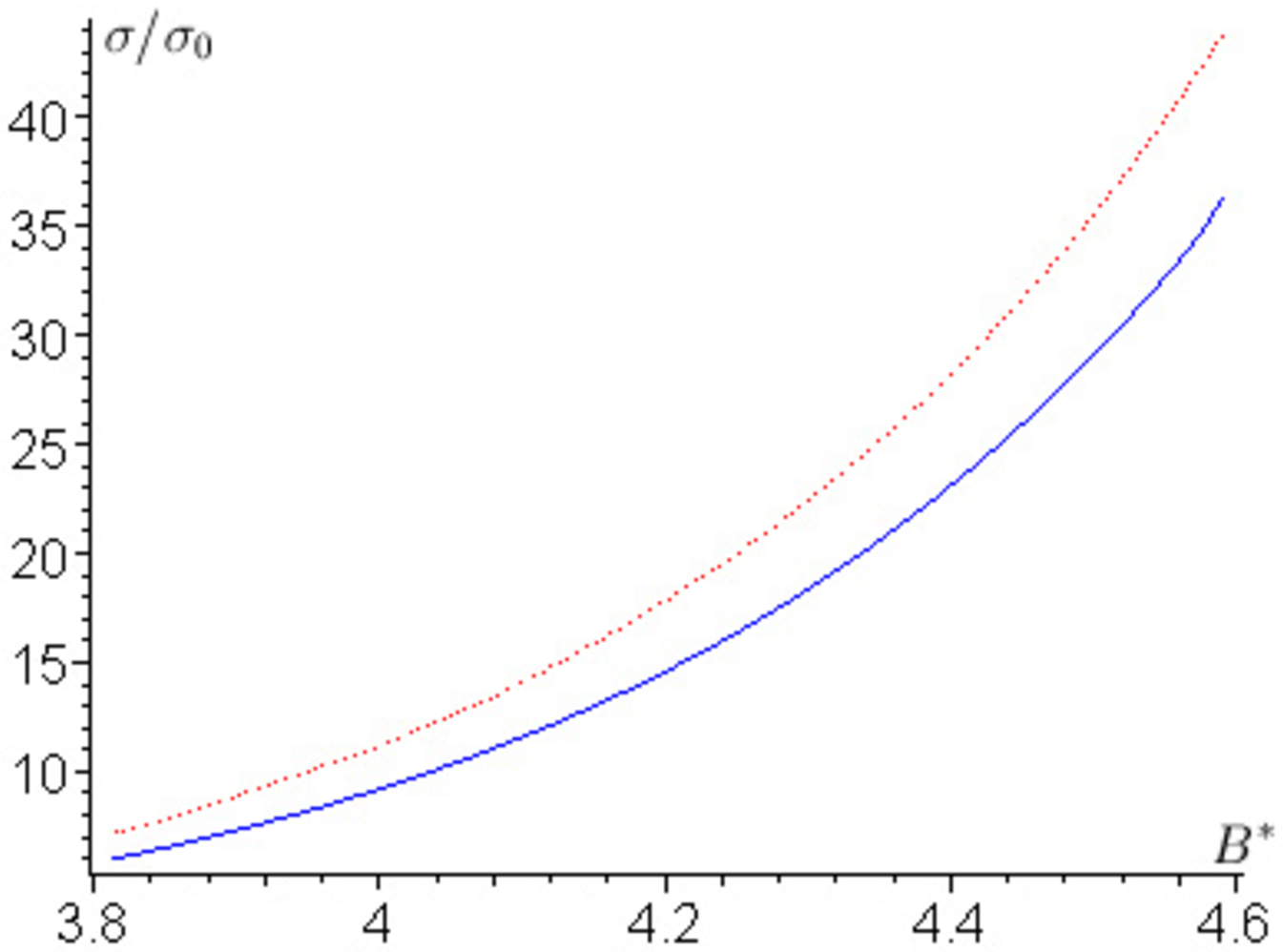}}
      \subfigure[]
  {\label{fig8c}
  \includegraphics[scale=0.55]{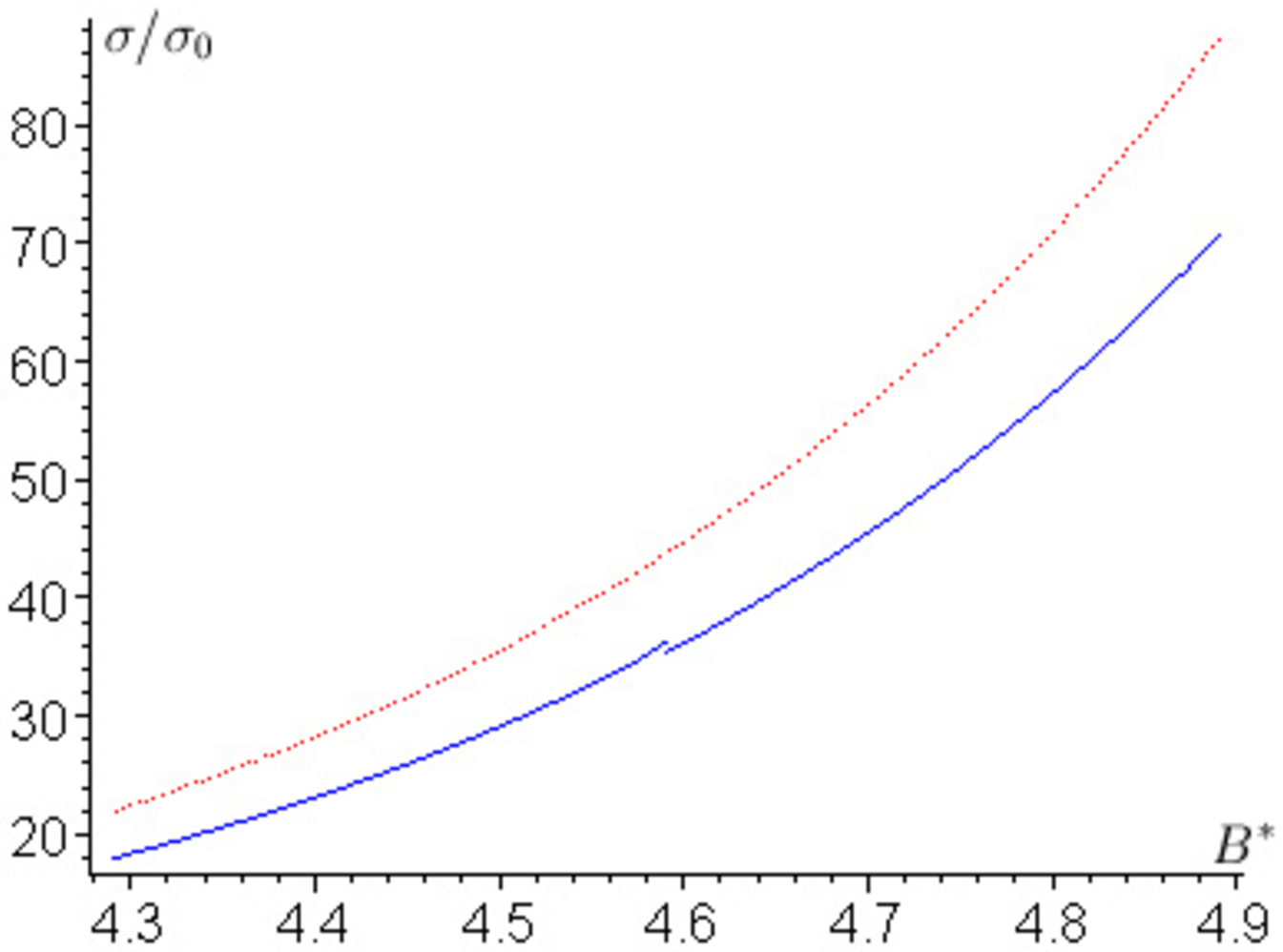}}
     \caption{\subref{fig8a}-\subref{fig8c} \small{The cross section  in the strong
magnetic field, normalized to the cross section in the field-free
case, for the different intervals within $B_{cr}\leq B <B'_{cr}$
in the case of unpolarized ($S=0$) neutrons. The neutrino energy
is equal to $\kap=10\ MeV$. The super-strong magnetic field $B\geq
B'_{cr} \ (B^{*}\sim 4.6)$ is also included in the panel (c). The
solid and dashed lines correspond to the initial neutrino
propagation along ($\cos \theta=1$) and against ($\cos \theta
=-1$) the magnetic field vector, respectively. The logarithmic
scale is used: $B^{*}=\log {{B \over B_{0}}}$, where $B_{0}={m^2
\over e}$.}}
\end{center}
\end{figure}
%%%%%%%%%%%%%%%%%%%%%%%%%%%%%%%%%%%%%%%%%%%%%%%%%%%%%%%%%%%%

%%%%%%%%%%%%%%%%%%%%%%%%%%%%%%%%%%%%%%%%%%%%%%%%%%%%%%%%%%%%
\begin{figure}[]
\begin{center}
  \subfigure[]
  {\label{fig9a}
    \includegraphics[scale=0.55]{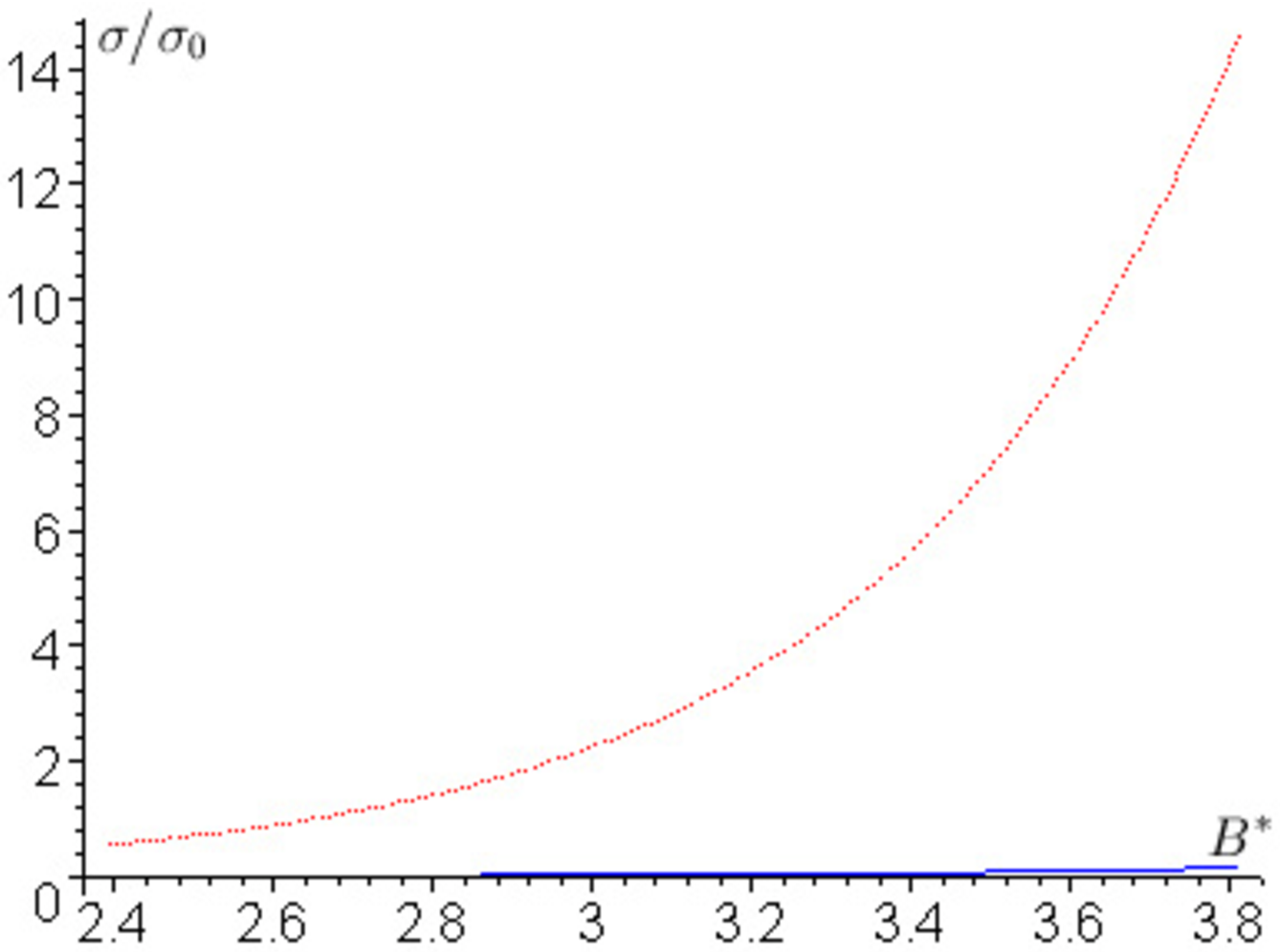}}
  %  \hspace{2cm}
  \subfigure[]
  {\label{fig9b}
  \includegraphics[scale=0.55]{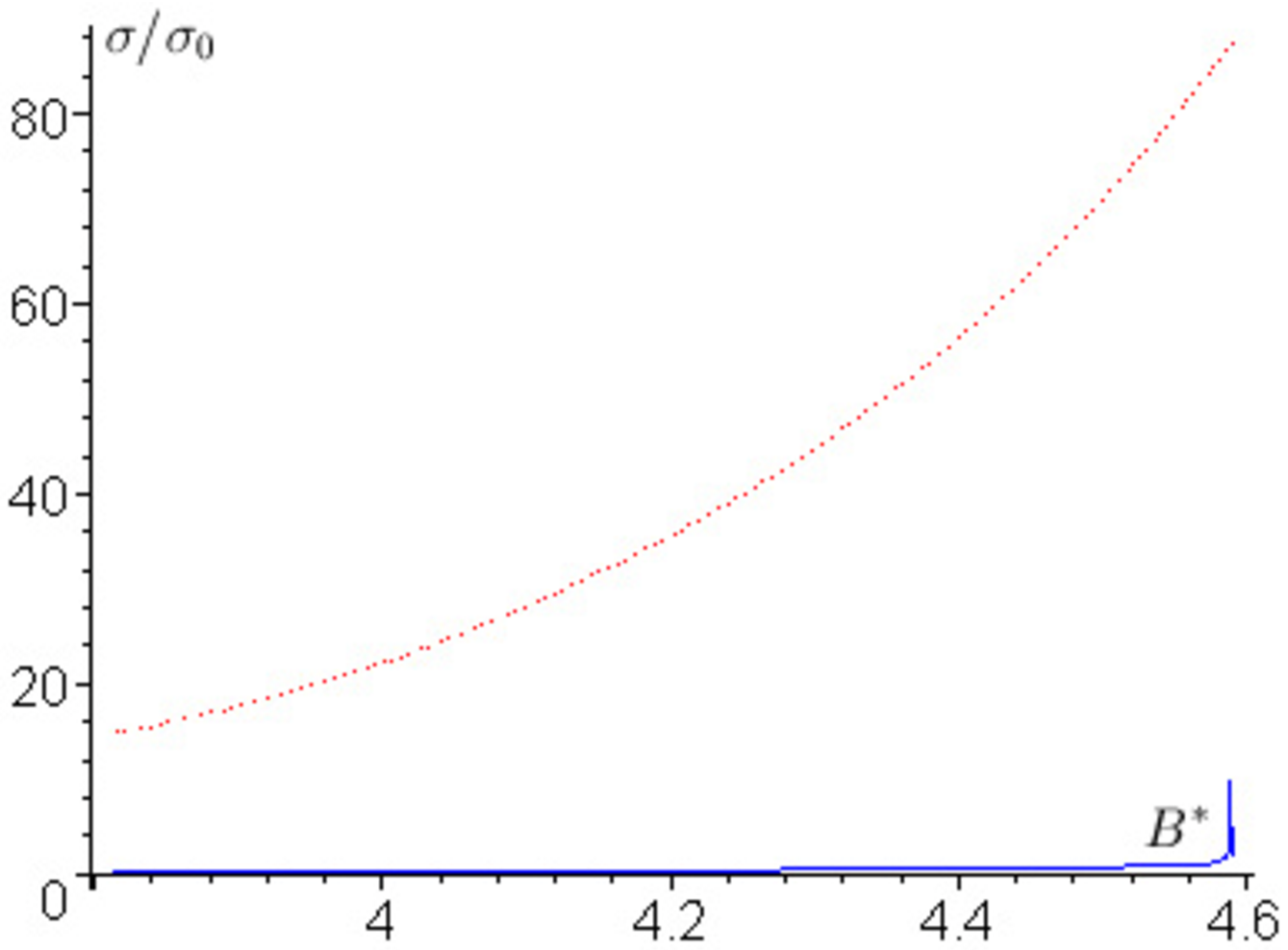}}
      \subfigure[]
  {\label{fig9c}
  \includegraphics[scale=0.55]{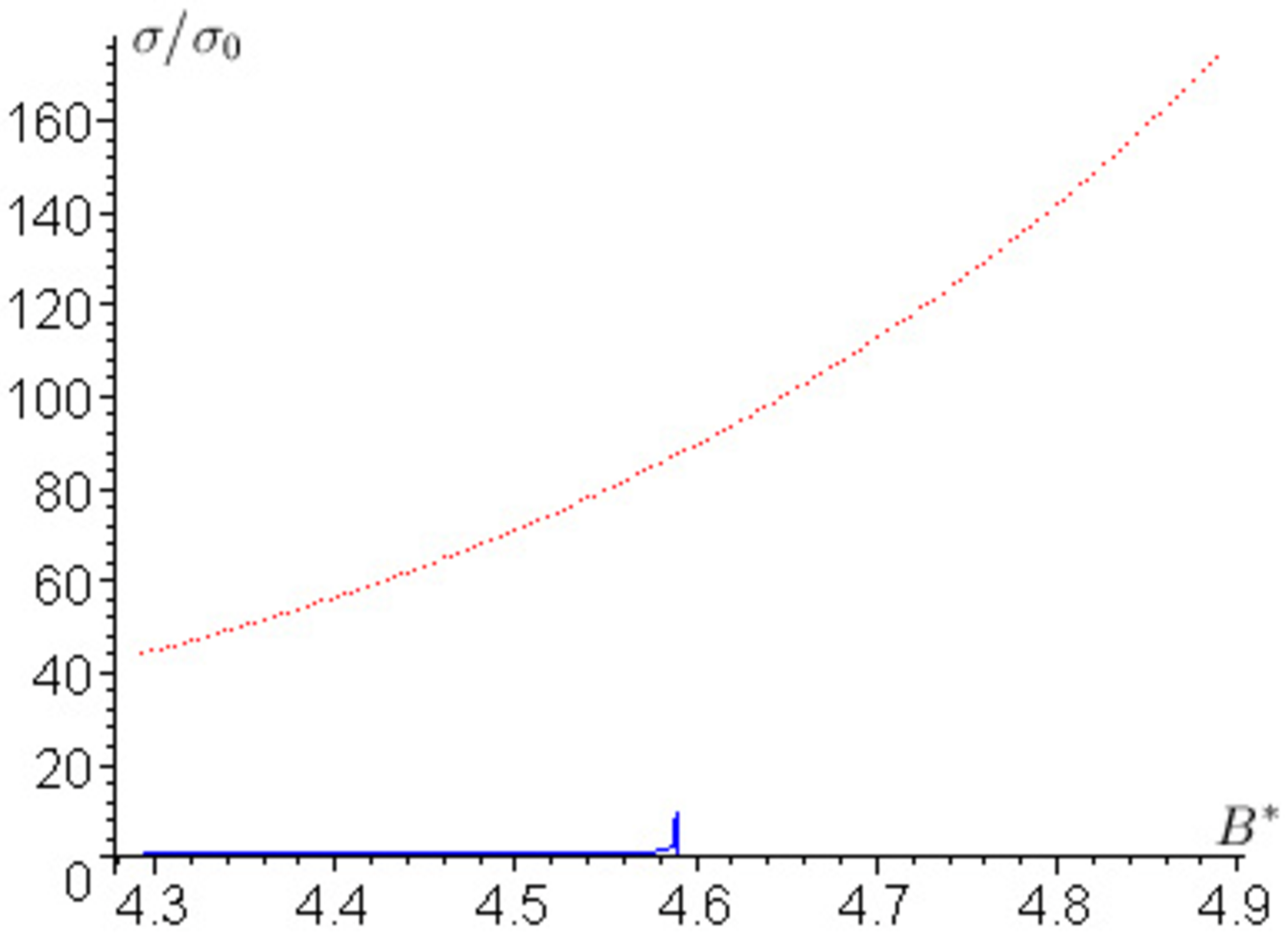}}
   % \hspace{2cm}
    \caption{\subref{fig9a}-\subref{fig9c}
    \small{The cross section  in the strong magnetic
field, normalized to the cross section in the field-free case, for
the different intervals within $B_{cr}\leq B < B'_{cr}$ in the
case of polarized neutrons with $S=-1$. The neutrino energy is
equal to $\kap=10\ MeV$. The super-strong magnetic field $B\geq
B'_{cr} \ (B^{*}\sim 4.6)$ is also included in the panel (c). The
solid and dashed lines correspond to the initial neutrino
propagation along ($\cos \theta=1$) and against ($\cos \theta
=-1$) the magnetic field vector, respectively. The cross section
in the case $\cos \theta =1$ is small for $B_{cr}\leq B < B'_{cr}$
and is zero for  $B\geq B'_{cr}$. The logarithmic scale is used:
$B^{*}=\log {{B \over B_{0}}}$, where $B_{0}={m^2 \over e}$.}}
\end{center}
\end{figure}

%%%%%%%%%%%%%%%%%%%%%%%%%%%%%%%%%%%%%%%%%%%%%%%%%%%%%%%%%%%%

%%%%%%%%%%%%%%%%%%%%%%%%%%%%%%%%%%%%%%%%%%%%%%%%%%%%%%%%%%%%
\begin{figure}[]
\begin{center}
  \subfigure[]
  {\label{fig10a}
    \includegraphics[scale=0.55]{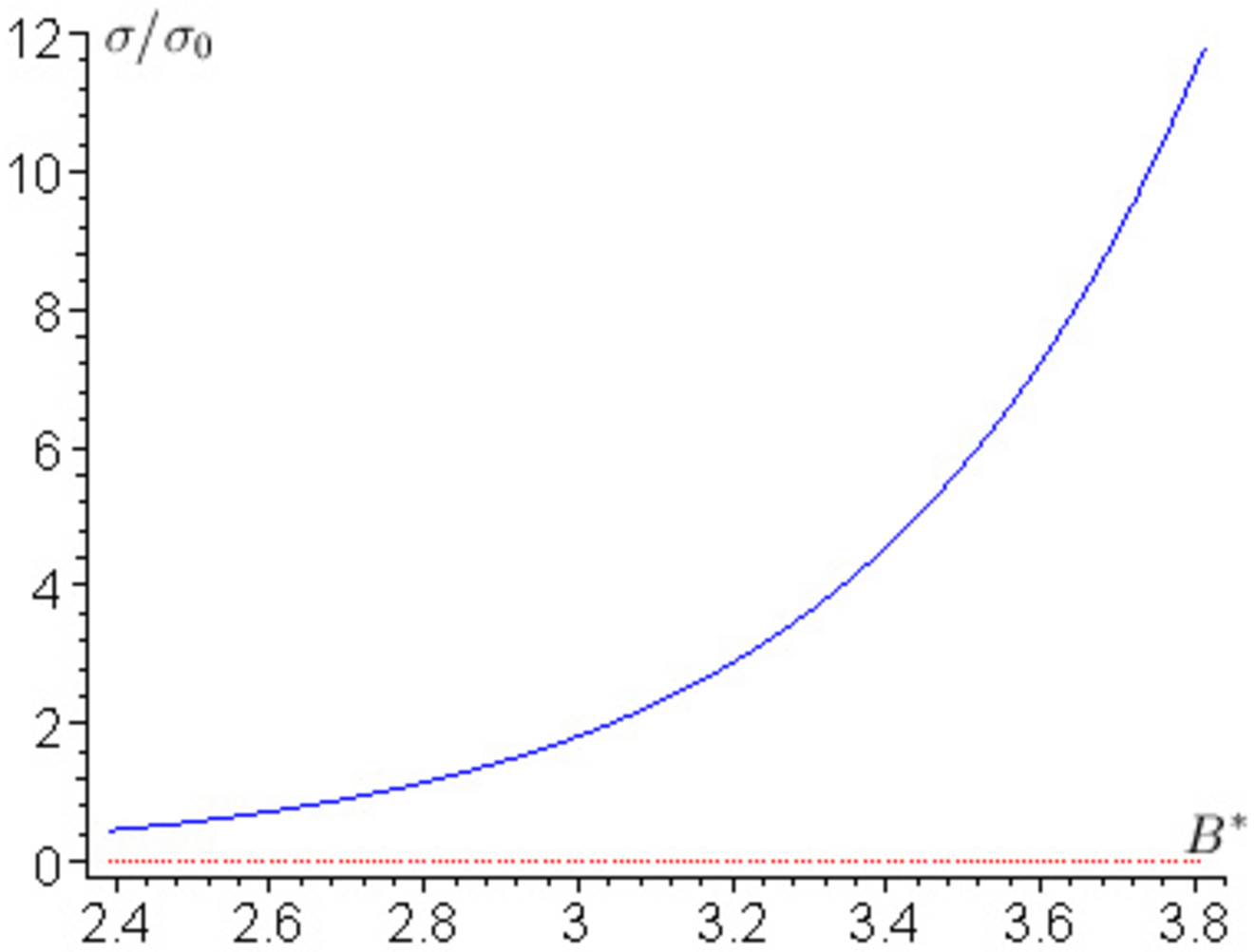}}
  %  \hspace{2cm}
  \subfigure[]
  {\label{fig10b}
  \includegraphics[scale=0.55]{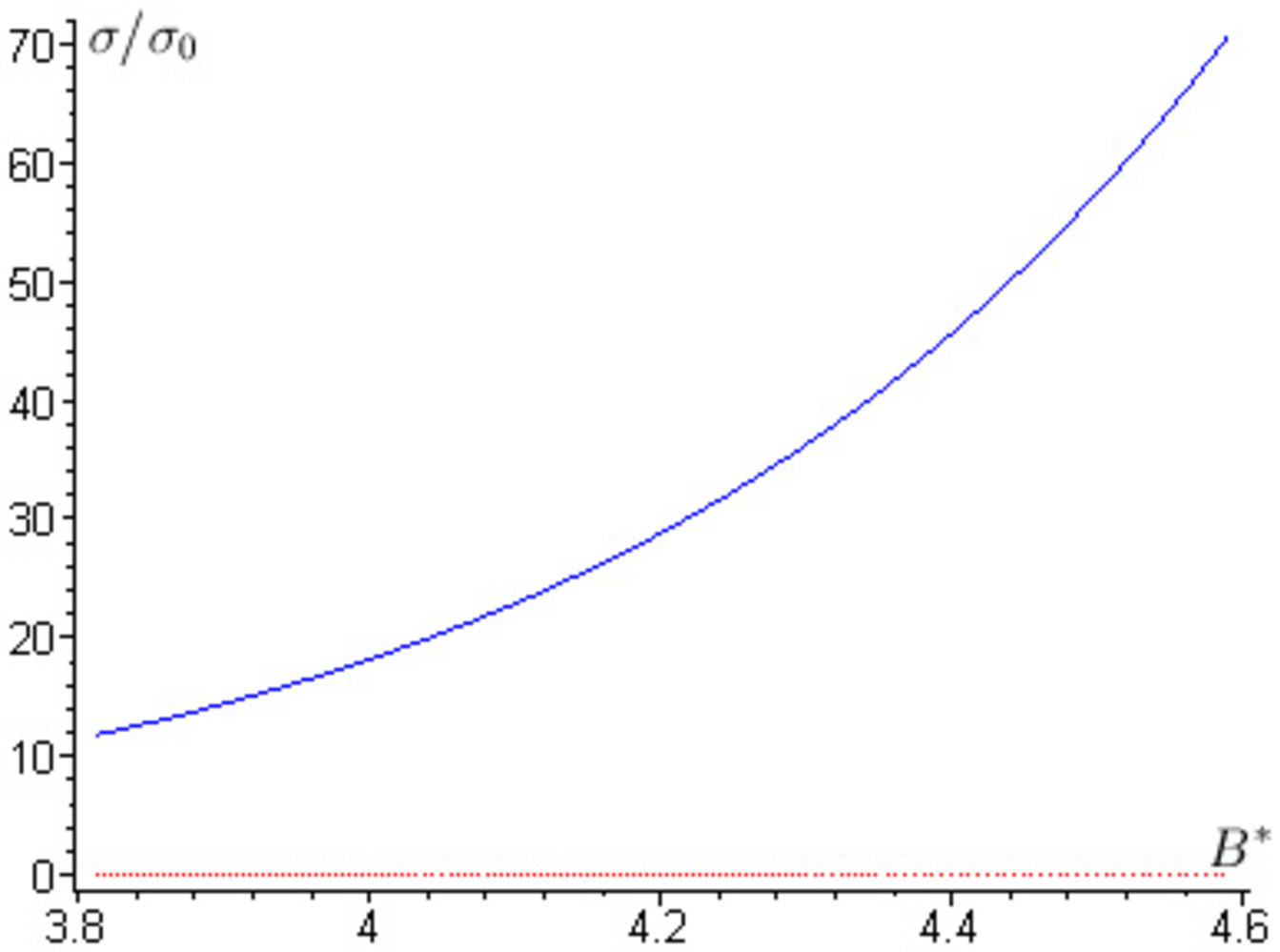}}
      \subfigure[]
  {\label{fig10c}
  \includegraphics[scale=0.55]{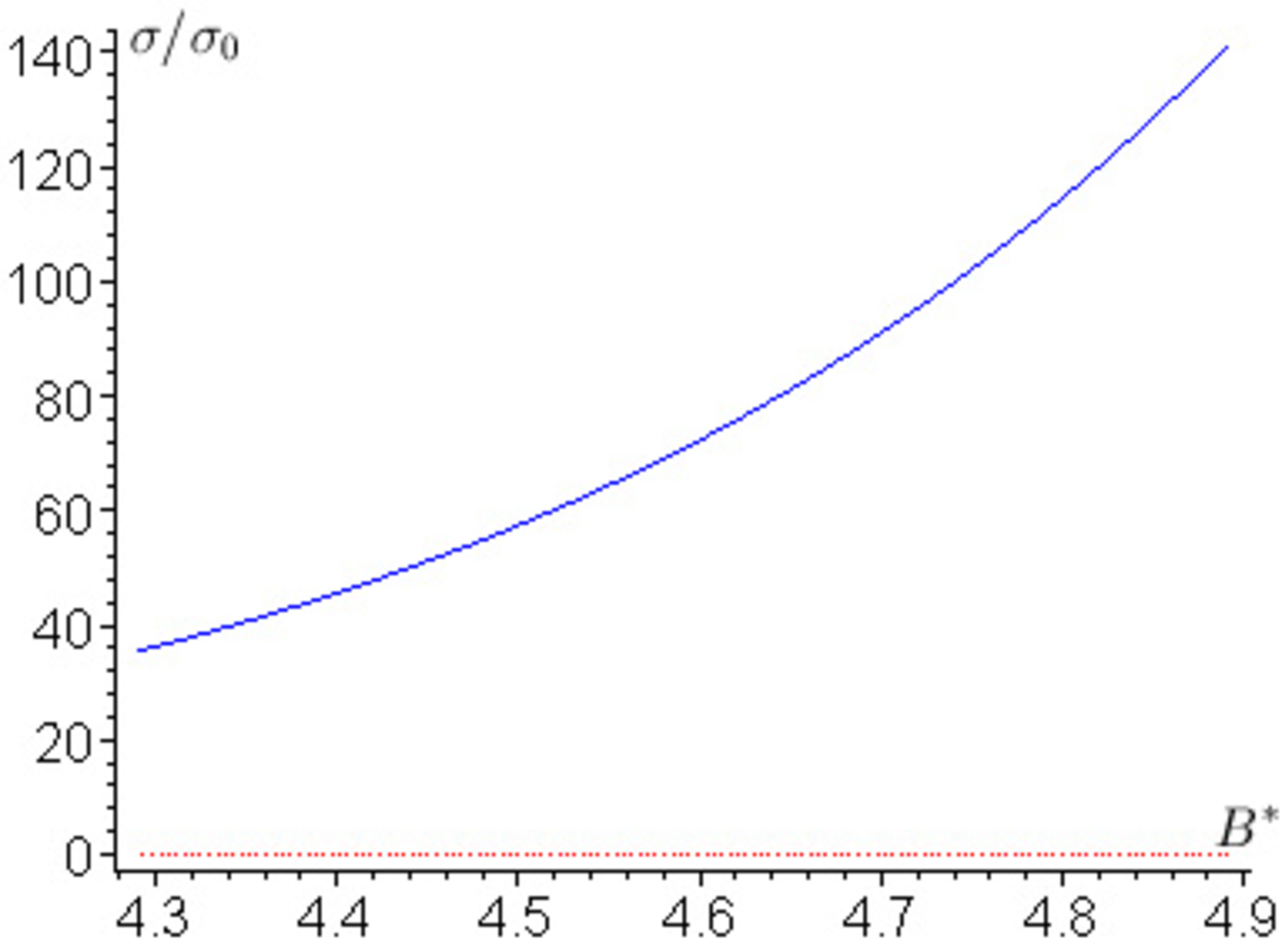}}
   % \hspace{2cm}
    \caption{\subref{fig9a}-\subref{fig10c} \small{The cross
    section  in the strong magnetic field, normalized to
    the cross section in the field-free case, for the different
intervals within $B_{cr}\leq B <B'_{cr}$ in the case of polarized
neutrons with $S=+1$.  The neutrino energy is equal to $\kap=10\
MeV$. The super-strong magnetic field $B\geq B'_{cr} \ (B^{*}\sim
4.6)$ also included in the panel (c). The solid and dashed lines
correspond to the initial neutrino propagation along ($\cos
\theta=1$) and against ($\cos \theta =-1$) the magnetic field
vector, respectively. The cross section for the case $\theta =-1$
is equal to zero. The logarithmic scale is used: $B^{*}=\log {{B
\over B_{0}}}$, where $B_{0}={m^2 \over e}$.}}
\end{center}
\end{figure}
%%%%%%%%%%%%%%%%%%%%%%%%%%%%%%%%%%%%%%%%%%%%%%%%%%%%%%%%%%%%

%%%%%%%%%%%%%%%%%%%%%%%%%%%%%%%%%%%%%%%%%%%%%%%%%%%%%%%%%%%%
\begin{figure}[h]
\begin{center}
  \subfigure[]
  {\label{fig11a}
    \includegraphics[scale=0.55]{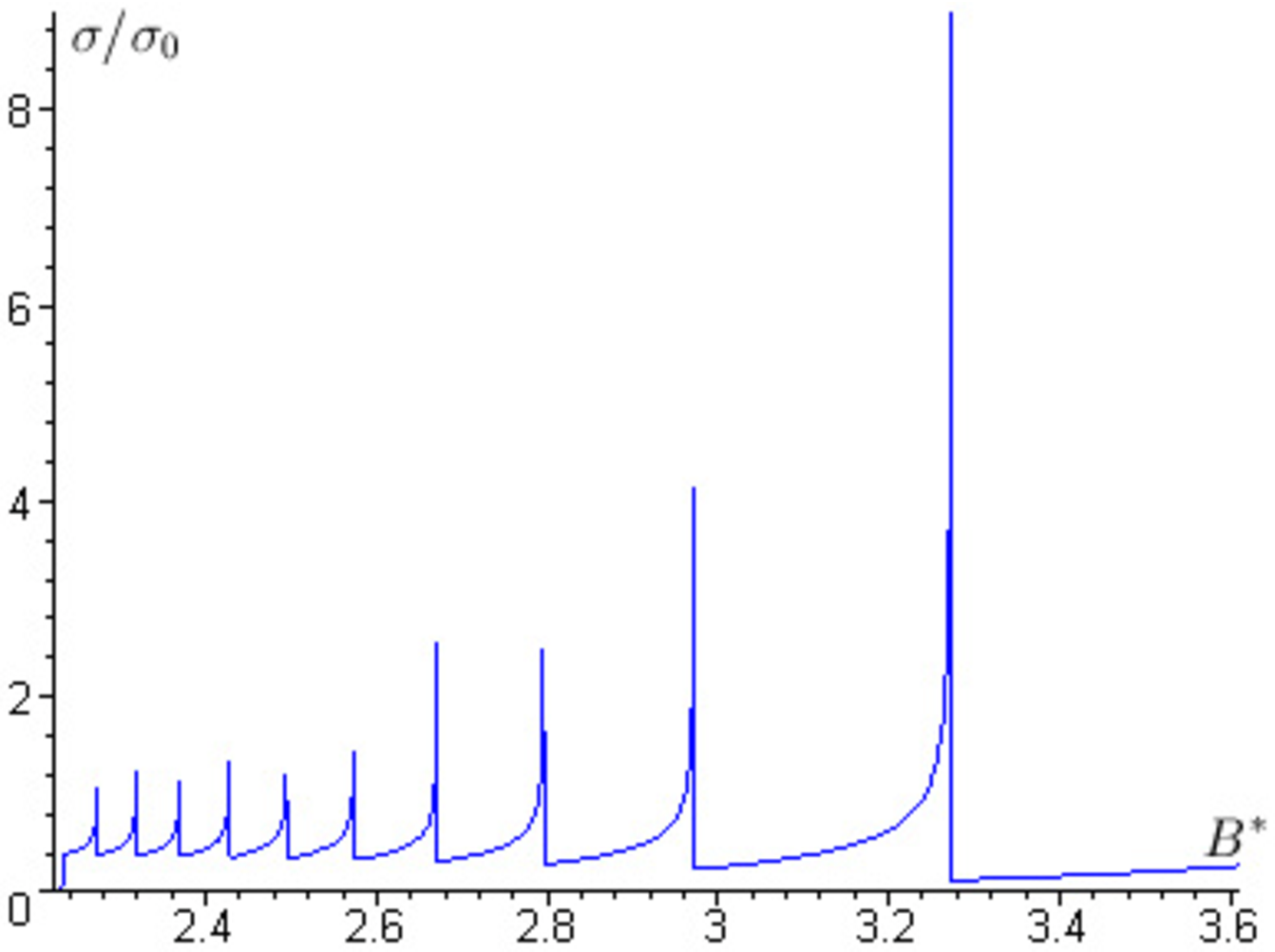}}
  %  \hspace{2cm}
  \subfigure[]
  {\label{fig11b}
  \includegraphics[scale=0.55]{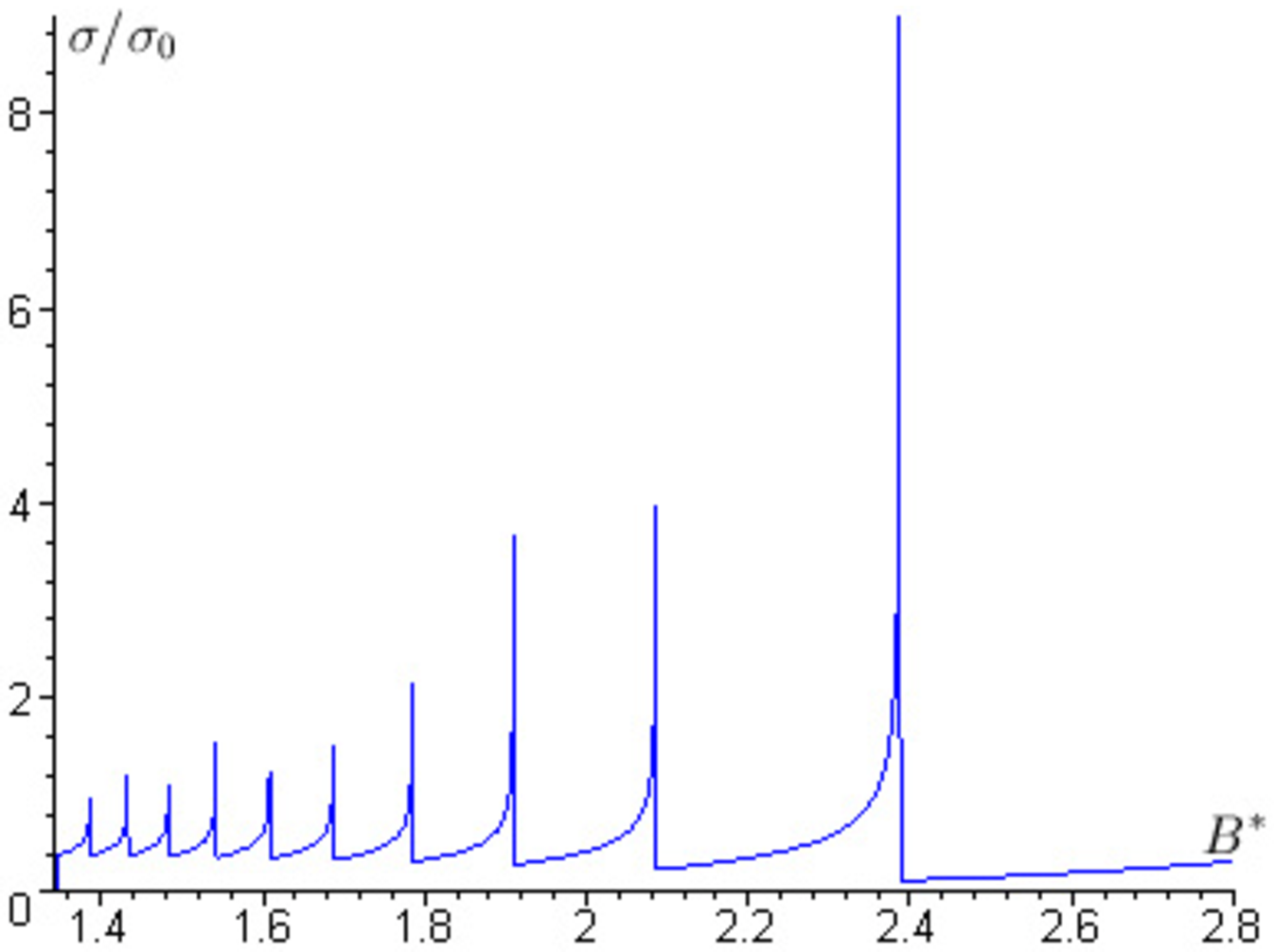}}
      \subfigure[]
  {\label{fig11c}
  \includegraphics[scale=0.55]{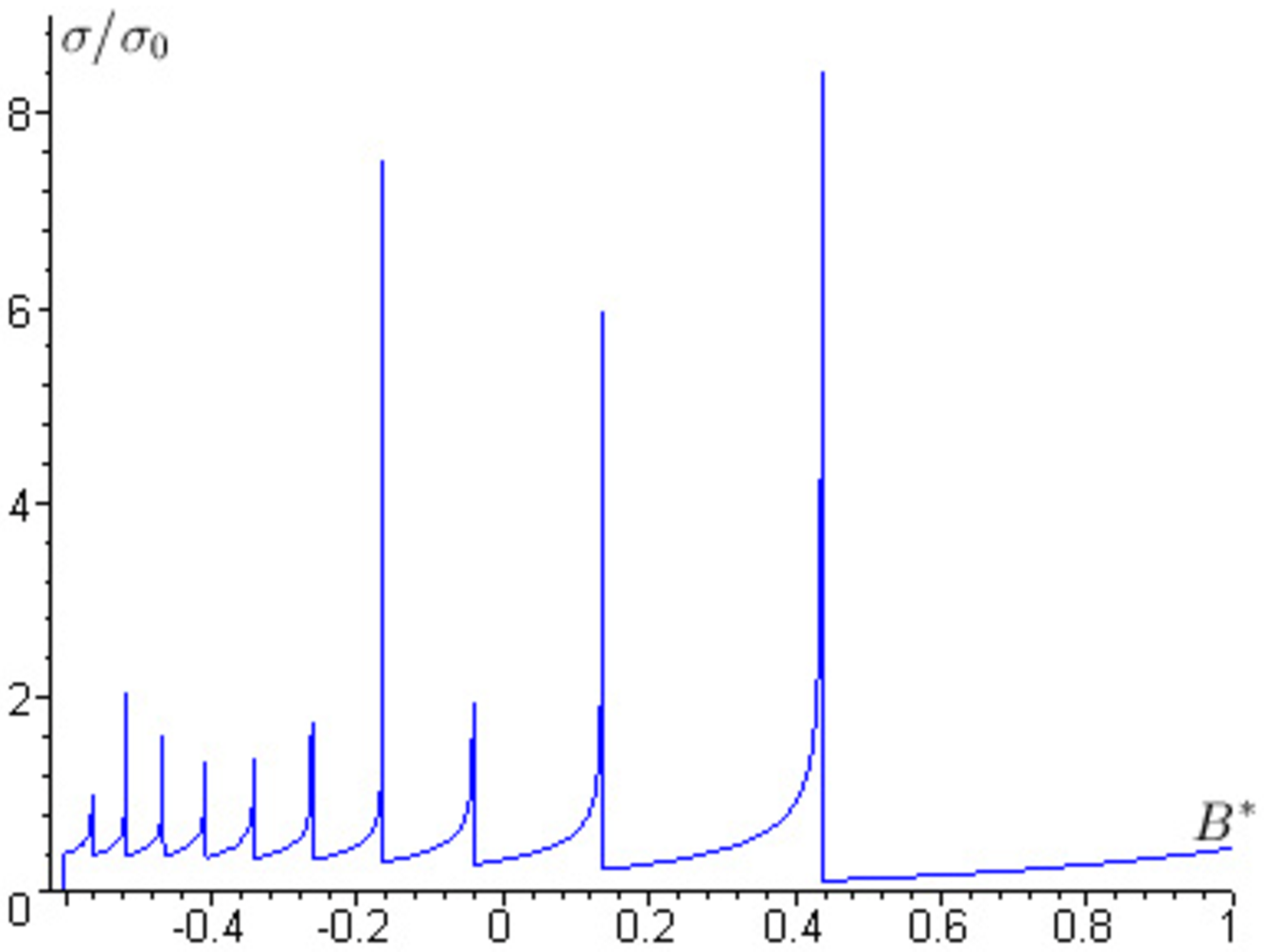}}
   % \hspace{2cm}
    \caption{\subref{fig11a}-\subref{fig11c}
    \small{The resonance behavior of the cross section in the
magnetic field, normalized to the field-free case,  for given
neutrino energies: $\kap=30 \ MeV$(a), $10\ MeV$ (b) and $\kap\ll
m$ (c). The logarithmic scale is used: $B^{*}=\log {{B \over
B_{0}}}$, where $B_{0}={m^2 \over e}$.}}
\end{center}
\end{figure}
%%%%%%%%%%%%%%%%%%%%%%%%%%%%%%%%%%%%%%%%%%%%%%%%%%%%%%%%%%%%

\end{document}